\documentclass[11pt]{article}
\usepackage{amssymb}
\usepackage{amsmath}
\usepackage{graphics}
\usepackage{epsfig}
\usepackage{a4wide}
\usepackage{cite}
\usepackage{multirow}
\usepackage{color}
\usepackage{mathrsfs}

\begin{document}

\title{ Double Higgs boson production and decay in Randall-Sundrum model at hadron colliders }
\author{ Zhang Wen-Juan$^a$, Ma Wen-Gan$^a$, Zhang Ren-You$^a$, Li Xiao-Zhou$^a$, Guo Lei$^b$, and Chen Chong$^a$ \\
{\small $^a$ Department of Modern Physics, University of Science and Technology of China, }  \\
{\small $~~$ Hefei, Anhui 230026, P.R. China} \\
{\small $^b$ Department of Physics, Chongqing University, Chongqing, 401331, P.R. China} }

\date{}
\maketitle \vskip 15mm

\begin{abstract}
We investigate the double Higgs production and decay at the $14~ {\rm TeV}$ LHC and $33~ {\rm TeV}$ HE-LHC in both the standard model and Randall-Sundrum (RS) model. In our calculation we consider reasonably only the contribution of the lightest two Kaluza-Klein (KK) gravitons. We present the integrated cross sections and some kinematic distributions in both models. Our results show that the RS effect in the vicinities of $M_{HH} \sim M_{1}$, $M_{2}$ (the masses of the lightest two KK gravitons) or in the central Higgs rapidity region is quite significant, and can be extracted from the heavy SM background by imposing proper kinematic cuts on final particles. We also study the dependence of the cross section on the RS model parameters, the first KK graviton mass $M_1$ and the effective coupling $c_0$, and find that the RS effect is reduced obviously with the increment of $M_1$ or decrement of $c_0$.
\end{abstract}
\vskip 3cm


\vfill \eject
\baselineskip=0.32in
\makeatletter      
\@addtoreset{equation}{section}
\makeatother       
\vskip 5mm
\renewcommand{\theequation}{\arabic{section}.\arabic{equation}}
\renewcommand{\thesection}{\Roman{section}.}
\newcommand{\nb}{\nonumber}

\section{Introduction}
\par
The huge gauge hierarchy between the Planck scale and electroweak (EW) scale in the standard model (SM) motivates the proposal of new physics beyond the SM. Among amounts of prospective candidates addressing the hierarchy problem, extra dimensions models stay distinctive for taking into consideration the gravity effects at TeV scale. There are two distinct mechanisms to eliminate the gauge scale disparity in higher dimensions scenario, the large extra dimensions (LED) model, also known as the Arkani-Hamed-Dimopoulos-Dvali (ADD) model \cite{1-ADD}, and the Randall-Sundrum (RS) model \cite{2-RS}.

\par
In the scenario of the ADD model, the spacetime is constituted of $D={4+n}$ dimensions. The graviton can travel in the $D$ dimensional bulk while the SM particles are constrained to the normal (3+1)-dimensional brane. The $D$-dimensional fundamental scale $M_D$ is related to the effective 4-dimensional Planck scale ${M_{Pl}}$ via ${M_{Pl}^2 \sim M_D^{n+2}R^n}$, where $R$ is the radius of the compactification torus of the extra $n$ dimensions. If $R$ is large enough, the fundamental scale can be around the EW scale (i.e., $M_D \sim {\rm TeV}$), therefore the gauge hierarchy problem is solved. However, the ADD model reintroduces a new hierarchy between the compactification scale $R^{-1} \sim {\rm eV-MeV}$ and the fundamental scale $M_D \sim {\rm TeV}$. In the scenario of the RS model, there is only one extra spacial dimension, which is compactified to an orbifold of the size of the order of $M_{Pl}^{-1}$. The spacetime in the RS model is warped and has a metric multiplied by an exponential factor which arises from the background $AdS_5$ spacetime. The gauge hierarchy problem is solved by the exponential factor and the new hierarchy in the ADD model is also avoided. Moreover, the distinct characteristic of the Kaluza-Klein (KK) graviton spectrum differing from that in the ADD model will lead to rich phenomenology at TeV colliders.

\par
The discovery of a new neutral boson with mass of $M_H \sim 126~ {\rm GeV}$, which is promising to be the SM Higgs boson, is announced by both the ATLAS and CMS collaborations at the CERN Large Hadron Collider (LHC) \cite{3-Higgs}. It is of great interest for physicists to probe the new found particle's properties to verify whether it is the SM Higgs boson or content of new physics. What's more, the measurement of couplings of Higgs boson with other particles as well as itself is desired to understand the mechanism of electroweak symmetry breaking. The double Higgs boson production provides an opportunity to probe the Higgs trilinear self-interaction and therefore reconstruct the Higgs potential.

\par
Up to now, the double Higgs boson production has been widely phenomenologically investigated at both $pp$ hadron colliders and $e^+e^-$ linear colliders \cite{HH-phe-1,HH-phe-2,HH-phe-3}. The double Higgs boson production at the LHC in the SM including the next-to-leading order (NLO) QCD corrections has been calculated in Ref.\cite{5-Higgs-nlo}. The evaluation for the next-to-next-to-leading order (NNLO) QCD corrections to SM Higgs boson pair production at hadron colliders within the large top-mass approximation can be found in Ref.\cite{4-Higgs-nnlo}. The Higgs boson pair production at hadron colliders beyond the SM has also been explored, such as in the SM with four generation quarks \cite{6-fourth}, littlest Higgs model \cite{7-lh}, universal extra dimensions model \cite{8-ued}, warped extra dimensions model \cite{warped-ED} and supersymmetry as well as LED model \cite{9-susy-led,12-sun,Mojtaba}. In addition, the Higgs boson pair production via vector boson fusion has been studied up to the QCD NNLO in the SM \cite{LingLS} and type-II two-Higgs-doublet model \cite{LiWH}.

\par
In this paper, we study the possible RS effect on double Higgs boson production at hadron colliders. The rest of this paper is organized as  follows: In section II we give a brief description of the RS model. The calculation setup of related subprocesses is presented in section III. In section IV we present the numerical results and discussion. Finally, we give a short summary in section V. The relevant Feynman rules are given in Appendix.

\vskip 5mm
\section{Related theory}
\par
The spacetime in the RS model is assumed to be a 5-dimensional bulk constituted of the (3+1)-dimensional Minkowski spacetime and a warped extra dimension which is compactified on an orbifold $S_1/Z _2$ with compactification radius $R_c$. At the fixed points $\phi=0$ and $\pi$ of the orbifold, two branes with opposite tensions, the UV brane (Planck brane) and the IR brane (${\rm TeV}$ brane), are set, respectively. It's assumed that the graviton propagates in the whole bulk while the SM particles are localized on the IR brane. By solving the corresponding 5-dimensional Einstein's field equation, we get a nonfactorizable metric of the bulk as
\begin{equation}
ds^2=e^{-2k R_c |\phi|}\eta_{\mu\nu}dx^\mu dx^\nu +{R_c}^2d\phi^2,
\end{equation}
where $0 \le |\phi| \le \pi$, $\eta_{\mu\nu}$ represents the ordinary Minkowski metric, and $k \sim {\cal O}(M_{Pl})$ is the curvature scale of $AdS_5$. The hierarchy between the Planck scale and EW scale is therefore generated by the exponential warp factor $e^{-k R_c \pi}$ via the relationship $M_{Pl} e^{-k R_c \pi} \sim {\cal O}({\rm TeV})$ requiring not too large $R_c$. To explore the gravity effects on the ${\rm TeV}$ brane at $\phi=\pi$, one can expand the graviton field, treated as the fluctuation around the background metric, into the RS KK modes $h_{\mu\nu}^{(n)}$ upon compactification. Then the effective 4-dimensional interaction Lagrangian of the RS model is given by \cite{10-kk}
\begin{equation}
\mathcal{L} = -\frac{1}{\overline{M}_{Pl}}T^{\mu\nu}h^{(0)}_{\mu\nu}-\frac{1}
              {\Lambda_{\pi}}T^{\mu\nu}\sum^{\infty}_{n=1}h^{(n)}_{\mu\nu},
\end{equation}
where $\Lambda_{\pi} = \overline{M}_{Pl} e^{-k R_c\pi}$, $\overline{M}_{Pl} = M_{Pl}/\sqrt{8\pi}$ is the reduced Planck scale, and $T^{\mu\nu}$ represents the SM energy-momentum tensor. The couplings of the zero mode ($n=0$) and massive modes ($n=1,2,...$) to SM particles are proportional to $1/\overline{M}_{Pl}$ and $1/\Lambda_{\pi}$, respectively. Due to the fact that
$\Lambda_{\pi}/\overline{M}_{Pl} \sim {\cal O}(10^{-16})$, the zero mode decouples from the graviton mass spectrum. The mass of the $n^{th}$ RS KK graviton can be written as
\begin{equation}
\label{G-KK-mass}
M_n = x_n k e^{-k R_c\pi}=\frac{x_n}{x_1}M_1,
\end{equation}
where $x_n$ is the $n^{th}$ root of the Bessel function, e.g., $x_1 \simeq 3.83$, $x_2 \simeq 7.02$ and $x_3 \simeq 10.17$. The mass splitting of the RS KK gravitons is of the ${\rm TeV}$ order, which implies that the RS KK gravitons can be produced as resonances at multi-TeV colliders.

\par
In this paper we choose the mass of the first KK mode $M_1$ and the effective coupling constant $c_0 \equiv k/\overline{M}_{Pl}$ as the two independent input parameters of the RS model. The relevant Feynman rules of RS KK gravitons' couplings to SM particles \cite{16-RS-coupling} are presented in Appendix. The effective graviton propagator, defined as a sum over infinite tower of RS KK gravitons, in the de Donder gauge can be expressed as
\begin{equation}\label{formula4}
G^{\mu\nu,\alpha\beta}_{KK} = \frac{1}{2} D(s) (\eta^{\mu\alpha} \eta^{\nu\beta} + \eta^{\mu\beta}\eta^{\nu\alpha} - \frac{2}{3}\eta^{\mu\nu} \eta^{\alpha\beta}),
\end{equation}
where
\begin{equation}\label{formula5}
D(s)=\sum^{\infty}_{n=1}\frac{i}{s-M_n^2+iM_n\Gamma_n},
\end{equation}
and $\Gamma_n$ is the total decay width of the $n^{th}$ KK graviton written as \cite{6-Total-width,22-Total-width}
\begin{equation} \label{formula5-1}
 \Gamma_n = {1 \over 16\pi}x_n^2M_nc_0^2\Delta_n ,
\end{equation}
with
\begin{equation}
\Delta_n = \Delta_n^{\gamma \gamma} + \Delta_n^{gg}
         + \Delta_n^{WW} + \Delta_n^{ZZ}
         + \sum_\nu \Delta_n^{\nu\nu} + \sum_l \Delta_n^{ll}
         + \sum_q \Delta_n^{qq}
         + \Delta_n^{HH} .
\end{equation}
$\Delta_n^{yy}$ is the coefficient for the decay $G^{(n)}_{KK} \to yy$, and $y$ is the SM particle involved. The explicit expressions for $\Delta_n^{yy}$ $(y=\gamma, g, W, Z, \nu, l, q, H)$ are given in Refs.\cite{6-Total-width, 21-Total-width}.

\vskip 5mm
\section{Calculation Setup}
\par
In our calculation we set the quark masses of the first two generations to zero, i.e., $m_{u}=m_{c}=m_{d}=m_{s}=0$, and consequently only consider top- and bottom-quark Yukawa couplings with Higgs boson. We use the dimensional regularization scheme in $D=4-2\epsilon$ dimensions to isolate UV and IR singularities and adopt the five-flavor scheme in the convolution with parton distribution functions (PDFs).

\par
\subsection{Double Higgs boson production in SM}
\label{Sec.III.1}
\par
Due to the masslessness of the first two generations of quarks, the dominant contribution to the Higgs boson pair production at a high energy hadron collider in the SM is from the $gg$ fusion and $b\bar{b}$ annihilation partonic processes. Although the lowest order amplitude squared for $gg \to HH$ is of the $\mathcal{O}(\alpha_{ew}^2\alpha_s^2)$, which is two orders of magnitude in $\alpha_s$ higher than that for $b\bar{b} \to HH$, the $gg$ fusion subprocess can be dominant channel at TeV-scale hadron colliders due to the high gluon luminosity. In this paper we consider the $gg \to HH$ and $b\bar{b} \to HH$ partonic processes only at the lowest order for the hadronic production of Higgs boson pair in the SM.

\par
{\bf A. Bottom-antibottom annihilation}
\par
\begin{figure}[!htbp]
\begin{center}
\includegraphics[scale=0.8]{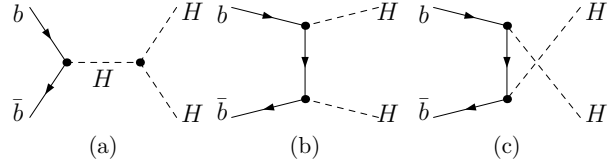}
\caption{\label{fig1} The tree-level Feynman diagrams for the $b \bar{b} \to HH$ partonic process in the SM.}
\end{center}
\end{figure}

The LO contribution from the $b\bar{b} \to HH$ channel at a high energy hadron collider in the SM is much less than from the $gg \to HH$ channel, e.g., the LO contributions from $b\bar{b} \to HH$ to total cross section at the $14~{\rm TeV}$ LHC and $33~{\rm TeV}$ HE-LHC are less than $0.185\%$ and $0.163\%$, respectively. Therefore, it is reasonable to include only the lowest order contribution for the $b\bar{b} \to HH$ channel in the SM calculation. The tree-level Feynman diagrams for the $b\bar{b}\to HH$ partonic process in the SM are shown in Fig.\ref{fig1}. The cross section for $b\bar{b}\to HH$ is expressed as
\begin{eqnarray}
\hat{\sigma}_{SM}^{b\bar{b}}(\hat{s})
=
\frac{1}{2} \frac{1}{4} \frac{1}{9}
\frac{(2 \pi)^4}{4|\vec{p}|\sqrt{\hat{s}}}
\int \sum_{{\rm spin}} \sum_{{\rm color}} |{\cal M}_{SM}^{0,b\bar{b}}|^2 d\Omega_2,
\end{eqnarray}
where $\vec{p}$ is the three-momentum of one of the incoming partons in center-of-mass system, ${\cal M}_{SM}^{0,b\bar{b}}$ is the Feynman amplitude for the tee-level diagrams in Fig.\ref{fig1}, and $d\Omega_2$ is the two-body phase space element. The first factor $\frac{1}{2}$ is due to the two identical Higgs bosons of final state, and the following two factors $\frac{1}{4}$ and $\frac{1}{9}$ are from the averaging of spins and colors of initial state.

\par
{\bf B. Gluon-gluon fusion}
\par
\begin{figure}[!htbp]
\begin{center}
\includegraphics[scale=0.8]{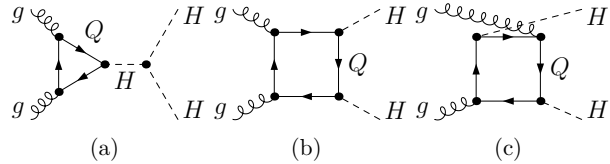}
 \caption{\label{fig2} The lowest-level Feynman diagrams for the $gg \to HH$ partonic process in the SM, where $Q$ represents massive quarks $t$ and $b$. (The diagrams with the exchange of initial and/or final identical particles are not shown.)}
\end{center}
\end{figure}

In Fig.\ref{fig2} we demonstrate some representative Feynman diagrams at the lowest order for the $gg \to HH$ partonic process in the SM. They are all one-loop graphs and the full one-loop amplitude for this partonic process, ${\cal M}_{SM}^{1,gg}$, is ultraviolet (UV) and infrared (IR) safe. The cross section for $gg \to HH$ at the lowest order, $\hat{\sigma}_{SM}^{gg}$, can be expressed as
\begin{eqnarray}\label{X-sec-gg-SM}
\hat{\sigma}_{SM}^{gg}(\hat{s})
=
\frac{1}{2} \frac{1}{4} \frac{1}{64}
\frac{(2 \pi)^4}{4|\vec{p}|\sqrt{\hat{s}}}
\int \sum_{{\rm spin}} \sum_{{\rm color}} |{\cal M}_{SM}^{1,gg}|^2 d\Omega_2.
\end{eqnarray}

\par
{\bf C. Integrated cross section}
\par
We denote $\sigma_{SM}^{b\bar{b}}$ and $\sigma_{SM}^{gg}$ as the integrated cross sections for $pp \to b\bar{b} \to HH + X$ and $pp \to gg \to HH + X$ in the SM, respectively. The two integrated cross sections can be obtained by integrating the parton-level cross sections, $\hat{\sigma}_{SM}^{b\bar{b}}$ and $\hat{\sigma}_{SM}^{gg}$, with the corresponding PDFs,
\begin{eqnarray} \label{pp-bbgg-total cross section}
&& \sigma_{SM}^{b\bar{b}}
= \int_0^1 dx_1 \int_0^1 dx_2 \Big[ G_{b/P_1}(x_1,\mu_f)G_{\bar{b}/P_2}(x_2,\mu_f) + (1 \leftrightarrow 2) \Big] \hat{\sigma}_{SM}^{b\bar{b}}(\hat s = x_1 x_2 s), \nonumber \\
&& \sigma_{SM}^{gg}
= \frac{1}{2} \int_0^1 dx_1 \int_0^1 dx_2 \Big[ G_{g/P_1}(x_1,\mu_f)G_{g/P_2}(x_2,\mu_f) + (1 \leftrightarrow 2) \Big] \hat{\sigma}_{SM}^{gg}(\hat s = x_1 x_2 s),~~~~~
\end{eqnarray}
where $G_{b,\bar{b},g/P}$ are the PDFs of bottom, antibottom and gluon in proton, $x_i~ (i=1,2)$ is the momentum fraction of a parton in proton $P_i$, and $\mu_f$ is the factorization scale. Then the integrated cross section for the parent process $pp \to HH + X$ in the SM is obtained as
\begin{eqnarray}
\label{full Xection-SM}
\sigma_{SM} = \sigma_{SM}^{gg} + \sigma_{SM}^{b\bar{b}}.
\end{eqnarray}

\par
\subsection{Double Higgs boson production in RS model}
\par
In the framework of the RS model, the Higgs boson pair can be produced via virtual KK graviton exchange, i.e.,
\begin{eqnarray}
pp \to gg/q\bar{q} \to G_{KK} \to HH + X,~~~~(q = u, d, c, s, b),
\end{eqnarray}
in addition to the production mechanism in the SM mentioned in section \ref{Sec.III.1}.

\par
{\bf A. Tree-level contribution}
\par
\begin{figure}[!htbp]
\begin{center}
\includegraphics[scale=0.8]{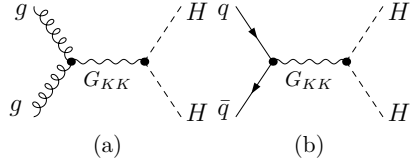}
\caption{\label{fig3} The tree-level Feynman diagrams for $gg/q\bar{q} \to G_{KK} \to HH$, where $q = u, d, c, s, b$, and $G_{KK}$ represents the KK graviton.}
\end{center}
\end{figure}

In the SM, only the $b\bar{b}$ annihilation can give tree-level contribution to the Higgs pair production at a hadron collider. However, in the RS model both the $gg$ fusion and all the $q\bar{q}$ $(q=u,d,c,s,b)$ annihilations can contribute at tree level to the Higgs pair production via KK graviton mediation, since the graviton field can couple with SM particles. The tree-level Feynman diagrams for $gg/q\bar{q} \to G_{KK} \to HH~(q = u, d, c, s, b)$ are presented in Fig.\ref{fig3}. The Feynman amplitudes for Figs.\ref{fig3}(a) and (b) are denoted as ${\cal M}_{KK}^{0,gg}$ and ${\cal M}_{KK}^{0,q\bar{q}}$, respectively. Therefore, the tree-level amplitudes for the $gg$ fusion and $q\bar{q}$ annihilation subprocesses in the RS model are
\begin{eqnarray}
&&{\cal M}_{RS}^{0,gg} = {\cal M}_{KK}^{0,gg}~~~~~~~~~~~~~~~~~~~~~~(gg~{\rm fusion}), \nonumber \\
&&{\cal M}_{RS}^{0,q\bar{q}} = {\cal M}_{KK}^{0,q\bar{q}}~~~~~~~~~~~~~~~~~~~~~~(q\bar{q}~{\rm annihilation},~q=u,d,c,s), \nonumber \\
&&{\cal M}_{RS}^{0,b\bar{b}} = {\cal M}_{SM}^{0,b\bar{b}} + {\cal M}_{KK}^{0,b\bar{b}}~~~~~~~~~~~(b\bar{b}~{\rm annihilation}).
\end{eqnarray}
We denote the tree-level cross sections for $gg \to HH$ and $q\bar{q} \to HH$ in the RS model as $\hat{\sigma}_{RS}^{0,gg}$ and $\hat{\sigma}_{RS}^{0,q\bar{q}}$, respectively.

\par
Since the loop contribution from the SM-like diagrams (shown in Fig.\ref{fig2}) for the $gg$ fusion subprocess is rather large, we include $\hat{\sigma}_{SM}^{gg}$ in the lowest order integrated cross section for the parent process $pp \to HH+X$ and therefore obtain
\begin{eqnarray}\label{LO-pp-cross-section}
\sigma_{RS}^0 &=& \Big[ \sigma_{SM}^{gg} + \sigma_{RS}^{0,gg} \Big] + \sum_{q=u,d}^{c,s,b} \sigma_{RS}^{0,q\bar{q}} \nb \\
&=&\frac{1}{2} \int_0^1 dx_1 \int_0^1 dx_2 \Big[ G_{g/P_1}(x_1,\mu_f)G_{g/P_2}(x_2,\mu_f) + (1 \leftrightarrow 2) \Big] \Big[ \hat{\sigma}_{SM}^{gg}(\hat s = x_1 x_2 s) + \hat{\sigma}_{RS}^{0,gg}(\hat s = x_1 x_2 s) \Big] \nb \\
&& + \sum_{q=u,d}^{c,s,b} \int_0^1 dx_1 \int_0^1 dx_2 \Big[ G_{q/P_1}(x_1,\mu_f)G_{\bar{q}/P_2}(x_2,\mu_f) + (1 \leftrightarrow 2) \Big] \hat{\sigma}_{RS}^{0,q\bar{q}}(\hat s = x_1 x_2 s).
\end{eqnarray}

\par
{\bf B. NLO QCD corrections }
\par
\begin{figure}[!htbp]
\begin{center}
\includegraphics[scale=0.8]{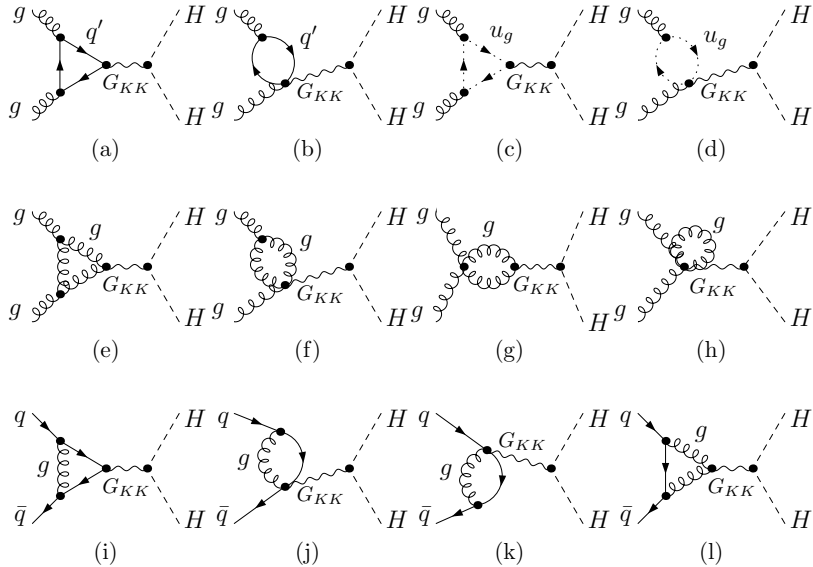}
 \caption{\label{fig4}
The QCD one-loop Feynman diagrams for $gg/q\overline{q} \to G_{KK} \to HH~(q=u,d,c,s)$, where $u_g$ is the ghost for gluon, $G_{KK}$ represents the KK graviton, and $q^{\prime}$ in fermion loops runs over $u, d, c, s, t, b$. (The diagrams with the exchange of initial two gluons are not shown.) }
\end{center}
\end{figure}

Due to the smallness of bottom-quark density in proton compared with gluon and light-quarks, the LO cross sections contributed by the $b\bar{b} \to HH$ subprocess at the LHC and HE-LHC in the RS model are less than $0.181\%$ and $0.153\%$, respectively. Therefore, we include only the LO contribution in the calculation of the $b\bar{b}\to HH$ subprocess in the RS model. Some representative QCD one-loop diagrams for $gg/q\bar{q} \to HH~ (q=u,d,c,s)$ involving KK graviton exchange are shown in Fig.\ref{fig4}. The full one-loop amplitudes for the $gg \to HH$ and $q\bar{q} \to HH~ (q=u,d,c,s)$ partonic processes in the RS model can be written as
\begin{eqnarray}
&&{\cal M}_{RS}^{1,gg} = {\cal M}_{SM}^{1,gg} + {\cal M}_{KK}^{1,gg}~~~~~~~~~~~(gg~{\rm fusion}), \nonumber \\
&&{\cal M}_{RS}^{1,q\bar{q}} = {\cal M}_{KK}^{1,q\bar{q}}~~~~~~~~~~~~~~~~~~~~~~~(q\bar{q}~{\rm annihilation},~q=u,d,c,s),
\end{eqnarray}
where ${\cal M}_{KK}^{1,gg}$ and ${\cal M}_{KK}^{1,q\bar{q}}$ are the one-loop amplitudes for $gg \to G_{KK} \to HH$ and $q\bar{q} \to G_{KK} \to HH$, respectively. In order to deal with the UV divergences, we introduce the quark and gluon wave-function renormalization constants as follows:
\begin{eqnarray}
q_{L,R}^{{\rm (bare)}}
=
\Big( 1 + \frac{1}{2} \delta Z^q_{L,R} \Big) q_{L,R}, ~~~~~~~
G_{\mu}^{a{\rm (bare)}}
=
\Big( 1 + \frac{1}{2} \delta Z^G \Big) G_{\mu}^a.
\end{eqnarray}
By adopting the on-shell renormalization scheme, these renormalization constants are fixed as
\begin{eqnarray}
&& \delta Z^q_{L,R}
=
-\frac{\alpha_{s}(\mu_{r})}{4\pi} C_F \biggl( \Delta_{UV} - \Delta_{IR} \biggr),~~~~~~~~~~~(q=u,d,c,s), \nonumber \\
&&~~ \delta Z^G
=
-\frac{\alpha_{s}(\mu_{r})}{4\pi}
 \left[
 \left( \frac{4}{3} n_f ^{UV} T_{F} - \frac{5}{3} C_{A} \right) \Delta_{UV}
-\left( \frac{4}{3} n_f ^{IR} T_{F} - \frac{5}{3} C_{A} \right) \Delta_{IR}
 \right] \nonumber \\
&& \,~~~~~~~~~~ -\frac{\alpha_{s}(\mu_{r})}{6\pi} \left[ \ln{\frac{\mu_{r}^2}{m_t^2}} + \ln{\frac{\mu_{r}^2}{m_b^2}} \right],
\end{eqnarray}
where $C_{F} = \frac{4}{3}$, $C_A = 3$, $T_{F} = \frac{1}{2}$, $\mu_r$ is the renormalization scale, $\Delta_{UV, IR} = \frac{1} {\epsilon_{UV, IR}} - \gamma_E + \ln4\pi$ are UV and IR regulators, $n_f^{UV} = 6$ corresponds to the six flavors of quarks, and $n_f^{IR} = 4$ is the number of massless quarks. After performing the renormalization procedure the UV singularities are removed, and therefore the NLO QCD virtual corrections to the $gg \to HH$ and $q\bar{q} \to HH~ (q=u,d,c,s)$ partonic processes in the RS model, $\hat{\sigma}_{RS}^{gg,V}$ and $\hat{\sigma}_{RS}^{q\bar{q},V}~ (q=u,d,c,s)$, are UV-finite.

\par
The renormalized virtual corrections are UV-finite, but still contain soft and collinear IR singularities, which can be canceled by adding the contributions of the related real emission processes and PDF counterterms. The real gluon emission partonic processes, $gg/q\bar{q} \to HHg~ (q=u,d,c,s)$, have both soft and collinear IR singularities which can be separated by applying the two cutoff phase space slicing (TCPSS) method \cite{13-tcpss}. Two cutoffs $\delta_s$ and $\delta_c$ are introduced in the TCPSS method to divide the phase space into soft ($S$), hard collinear ($HC$) and hard noncollinear ($\overline{HC}$) regions. Then the cross sections for the real gluon emission partonic processes can be expressed as
\begin{eqnarray}
\hat{\sigma}_{RS}^{ab,R} = \hat{\sigma}_{RS}^{ab,S} + \hat{\sigma}_{RS}^{ab,HC} + \hat{\sigma}_{RS}^{ab,\overline{HC}},
\end{eqnarray}
where $ab=gg, q\bar{q}$ correspond to the $gg$ fusion and $q\bar{q}$ annihilation, respectively. The hard noncollinear cross section $\hat{\sigma}_{RS}^{ab,\overline{HC}}$ is IR-finite, and the soft IR singularity in $\hat{\sigma}_{RS}^{ab,S}$ can be canceled exactly by that in the virtual correction $\hat{\sigma}_{RS}^{ab,V}$ as demonstrated by the Kinoshita-Lee-Nauenberg theorem \cite{KLN}. The collinear IR singularity in $\hat{\sigma}_{RS}^{ab,HC}$ is partially canceled by that in the virtual correction $\hat{\sigma}_{RS}^{ab,V}$, and the remaining collinear divergence can be absorbed by the corresponding PDF counterterms. Then the full NLO QCD corrections to the $gg \to HH$ and $q\bar{q} \to HH~ (q = u,d,c,s)$ partonic processes in the RS model are obtained as
\begin{eqnarray}
\label{NLO gg-qq-corrections}
\Delta \hat{\sigma}_{RS}^{ab} = \hat{\sigma}_{RS}^{ab,V} + \hat{\sigma}_{RS}^{ab,R} + \hat{\sigma}_{RS}^{ab,PDF},~~~~~~~(ab = gg, q\bar{q}),
\end{eqnarray}
where $q = u,d,c,s$ and the superscripts $V$, $R$, $PDF$ represent the virtual, real and PDF counterterm contributions, respectively.

\par
The real light-quark emission partonic processes, $gq \to HHq$ and $g\bar{q} \to HH\bar{q}~ (q = u,d,c,s)$, contain only collinear IR singularities. We separate the phase space into collinear ($C$) and noncollinear ($\overline{C}$) regions by using the cutoff $\delta_c$. The collinear IR singularities in the real light-quark emissions are also canceled by the PDF counterterms. Then we obtain the corrections from the real light-quark emissions and corresponding PDF counterterms as \footnote{Due to the CP conservation, the real emission correction and corresponding PDF counterterm contribution for $gq \to HH q$ are the same as for $g\bar{q} \to HH \bar{q}$, respectively, i.e., $\hat{\sigma}_{RS}^{gq,R} = \hat{\sigma}_{RS}^{g\bar{q},R}$ and $\hat{\sigma}_{RS}^{gq,PDF} = \hat{\sigma}_{RS}^{g\bar{q},PDF}$.}
\begin{eqnarray}
\label{NLO gq-gqbar-corrections}
&& \Delta \hat{\sigma}_{RS}^{gq} = \hat{\sigma}_{RS}^{gq,R} + \hat{\sigma}_{RS}^{gq,PDF},~~~
\Delta \hat{\sigma}_{RS}^{g\bar{q}} = \hat{\sigma}_{RS}^{g\bar{q},R} + \hat{\sigma}_{RS}^{g\bar{q},PDF},~~~~~~~(q = u,d,c,s).
\end{eqnarray}
The PDF counterterm contributions in Eqs.(\ref{NLO gg-qq-corrections}) and (\ref{NLO gq-gqbar-corrections}) can be expressed as
\begin{eqnarray}
&& \hat{\sigma}_{RS}^{gg,PDF} = 2 P_{gg} \otimes \hat{\sigma}_{RS}^{0,gg}, ~~~
\hat{\sigma}_{RS}^{q\bar{q},PDF} = 2 P_{qq} \otimes \hat{\sigma}_{RS}^{0,q\bar{q}}, \nb \\
&& \hat{\sigma}_{RS}^{gq,PDF} = \hat{\sigma}_{RS}^{g\bar{q},PDF} = P_{gq} \otimes \hat{\sigma}_{RS}^{0,gg} + P_{qg} \otimes \hat{\sigma}_{RS}^{0,q\bar{q}}.
\end{eqnarray}
$P_{gg}$, $P_{qq}$, $P_{gq}$ and $P_{qg}$ are the QCD splitting functions \cite{13-tcpss},
\begin{eqnarray}
&& P_{gg}(z) = 6 \left[ \frac{z}{1-z} + \frac{1-z}{z} + z(1-z) \right],~~~P_{qq}(z) = C_F \frac{1+z^2}{1-z},  \nb \\
&& P_{gq}(z) = C_F \frac{1+(1-z)^2}{z},~~~P_{qg}(z) = T_F \Big[ z^2 + (1-z)^2 \Big],
\end{eqnarray}
and the ``$\otimes$-convolution'' is defined as
\begin{eqnarray}
\Big[ P \otimes \hat{\sigma} \Big](\hat{s})
= \frac{1}{\epsilon}
\left[
\frac{\alpha_s}{2 \pi} \frac{\Gamma(1 - \epsilon)}{\Gamma(1 - 2 \epsilon)}
\left(
\frac{4 \pi \mu_r^2}{\mu_f^2}
\right)^{\epsilon}
\right]
\int_0^1 dz P(z) \hat{\sigma}(z\hat{s}).
\end{eqnarray}

\par
{\bf C. Integrated cross section}
\par
Due to the high gluon luminosity and low bottom-quark density in proton at the LHC and HE-LHC, we include the ${\cal O}(\alpha_{ew}^2\alpha_s^2)$ SM-like contribution (see Eq.(\ref{X-sec-gg-SM})) to the $gg$ fusion and only consider the tree-level contribution to the $b\bar{b}$ annihilation. Then the integrated cross section for the double Higgs boson production at a $pp$ collider in the RS model can be written as
\begin{eqnarray}
\label{full Xection-RS-nlo}
\sigma_{RS}=\sigma_{RS}^0 + \Delta \sigma_{RS},
\end{eqnarray}
where $\Delta \sigma_{RS}$ is the full NLO QCD correction to the $pp \to HH + X$ process obtained by convoluting the parton-level corrections, $\Delta \hat{\sigma}_{RS}^{gg}$, $\Delta \hat{\sigma}_{RS}^{q\bar{q}}$, $\Delta \hat{\sigma}_{RS}^{gq}$ and $\Delta \hat{\sigma}_{RS}^{g\bar{q}}$, with the corresponding PDFs,
\begin{eqnarray}
\label{full correction to parent process}
\Delta \sigma_{RS}=\sum_{ab \in {\cal S}}
\frac{1}{1+\delta_{ab}} \int_0^1 dx_1 \int_0^1 dx_2 \Big[ G_{a/P_1}(x_1,\mu_f) G_{b/P_2}(x_2,\mu_f) + (1 \leftrightarrow 2) \Big] \Delta \hat{\sigma}_{RS}^{ab}(\hat{s} = x_1 x_2 s), \nonumber \\
{\cal S} =\left\{ gg, \, q\bar{q}, \, gq, \, g\bar{q}\, | \, (q = u,d,c,s) \right\}.~~~~~~~
\end{eqnarray}

\vskip 5mm
\section{Numerical results and discussion }
\par
\subsection{Input parameters }
\par
The SM input parameters used in our calculation are taken as \cite{18-SMpara,CMS-2013}
\begin{eqnarray}
&& \alpha_{ew}(0) = 1/137.035999074, ~~m_t = 173.21~{\rm GeV}, ~~m_b = 4.75~{\rm GeV}, \nonumber \\
&& M_W = 80.385~{\rm GeV}, ~~~M_Z = 91.1876~{\rm GeV}, ~~~M_H = 126~{\rm GeV}.
\label{SMpar}
\end{eqnarray}
We checked numerically the independence of the integrated cross section on the two cutoffs $\delta_s$ and $\delta_c$, and take $\delta_s=10^{-4}$ and $\delta_c=\delta_s/50$ in further numerical calculation. We adopt the MSTW2008nlo PDF set \cite{14-pdf} with $n_f = 5$ and $\alpha_s(M_Z) = 0.12018$, and set the factorization and renormalization scales being equal for simplicity, i.e., $\mu_f = \mu_r = \mu$. To estimate the theoretical uncertainty from the factorization/renormalization scale, we investigate the scale dependence of the integrated cross sections in both the SM and RS model. In this paper, we take $\mu_0 = M_H$ and $\mu_1 = M_{HH}$ (Higgs pair invariant mass) as two typical central scales.

\par
The theoretical constraint on the effective coupling $c_0$ in the RS model is $c_0 \in [0.01,0.1]$ \cite{10-kk}. Up to now, no signature of the RS model has been observed and all the experimental data are in good agreement with the SM predictions, which gives more stringent constraints on the RS model parameters. Recently a lower bound on the first KK graviton mass was given by the ATLAS collaboration at $95\%$ confidence level as $M_1 > 1.23$ and $2.68~ {\rm TeV}$ for $c_0 = 0.01$ and $0.1$, respectively \cite{15-M1}. In the following numerical calculation we take $M_{1}=2.75~{\rm TeV}$ and $c_0 = 0.1$ as the RS model input parameters. Then we obtain the mass of the second KK graviton as $M_2=5.04~{\rm TeV}$ from Eq.(\ref{G-KK-mass}).

\par
The effective graviton propagator is a sum over the infinite tower of KK gravitons. Since the KK graviton mass is proportional to the root of Bessel function which increases notably, we may apply a cut on the number of active KK gravitons instead of taking the infinite tower of KK gravitons into consideration. We calculate the cross section $\sigma_{RS}^0$ for the Higgs pair production at the $14~{\rm TeV}$ LHC and $33~{\rm TeV}$ HE-LHC with different numbers of active KK gravitons, and find that the contribution of the $n^{{\rm th}}$ $(n>2)$ KK gravitons is less than $0.5\%$ and can be neglected. Therefore, in the following calculation we consider only the dominant contribution of the first two KK gravitons.

\par
\subsection{Integrated cross section }
\par
In Table \ref{tab1} we list the integrated cross sections for $pp \to HH+X$ in the SM and RS model, $\sigma_{SM}$ and $\sigma_{RS}$, at the $14~{\rm TeV}$ LHC and $33~{\rm TeV}$ HE-LHC. In order to describe the RS effect quantitatively, we define the relative RS effect as $\delta = \left(\sigma_{RS}-\sigma_{SM}\right)/\sigma_{SM}$. From the table we can see that the relative RS effect is about $3.15\%$ at the $14~ {\rm TeV}$ LHC, and can reach $14.75\%$ at the $33~ {\rm TeV}$ HE-LHC which could be detectable in experiment.
\begin{table}[!htp]
\begin{center}
\begin{tabular}{c|c|c|c}
\hline
~$\sqrt{S}$ [TeV]~ & ~~$\sigma_{SM}$ $[fb]$~~ & ~~$\sigma_{RS}$ $[fb]$~~ & ~~$\delta$ $[\%]$~~ \\
\hline \hline
14 & 14.8703(9) & 15.339(1) & 3.15 \\
33 & 91.509(7) & 105.01(1) & 14.75\\
\hline
\end{tabular}
\end{center}
\caption{\label{tab1} The integrated cross sections for $pp \to HH+X$ in the SM and RS model ($\sigma_{SM}$ and $\sigma_{RS}$) and the corresponding relative RS effects at the $14~{\rm TeV}$ LHC and $33~{\rm TeV}$ HE-LHC with $\mu=\mu_1$. }
\end{table}

\par
The RS effect is mainly contributed by the KK graviton resonance production when partonic colliding energy is greater than KK graviton mass. In order to enhance the RS effect we apply a lower cut on the invariant mass of final Higgs boson pair as shown in Table \ref{tab2}. In this table we provide the integrated cross sections for $pp \to HH + X$ in both the SM and RS model and the corresponding relative RS effects with the constraint of $M_{HH}> M_{HH}^{cut}$ at the $14~ {\rm TeV}$ LHC and $33~ {\rm TeV}$ HE-LHC. We can see clearly that both $\sigma_{SM}$ and $\sigma_{RS}$ decrease, but the relative RS effect increases, with the increment of the Higgs pair invariant mass cut $M_{HH}^{cut}$. For example, the relative RS effects are $3.21\%$ and $14.92\%$ at the $14~ {\rm TeV}$ LHC and $33~ {\rm TeV}$ HE-LHC with the constraint of $M_{HH} > 300~{\rm GeV}$, and increase to $336\%$ and $791\%$, respectively, after applying the cut of $M_{HH} > 1000~{\rm GeV}$.

\par
In Table \ref{tab2} we list both $\sigma_{RS}^0$ and $\sigma_{RS}$ in the RS model to demonstrate the effect of the NLO QCD correction. We see that the NLO QCD correction always enhances the LO cross section in the RS model, therefore neglecting the NLO QCD correction would lead to an underestimation of the RS effect. With the increment of $M_{HH}^{cut}$, the NLO QCD correction in the RS model becomes more and more significant, the production rates for $pp \to HH + X$ in the SM and RS model go down quickly while the relative RS effect increases rapidly. We can read out from the table that the NLO QCD corrected cross section and the relative RS effect are $0.601~fb$, $336\%$ at the $14~ {\rm TeV}$ LHC, and $15.15~fb$, $791\%$ at the $33~ {\rm TeV}$ HE-LHC, respectively, with the constraint of $M_{HH} > 1000~{\rm GeV}$. In the following calculation we fix the lower cut on the Higgs pair invariant mass as $M_{HH}^{cut} =1000~{\rm GeV}$.

\par
In the SM the dominant channel for Higgs pair production at hadron colliders is gluon-gluon fusion via virtual top quark. This production channel begins at one-loop level, however, the higher order QCD corrections to this process are quite significant. The QCD corrections up to the NNLO to the SM Higgs pair production at hadron colliders within the large top-mass approximation are already available in Ref.\cite{4-Higgs-nnlo}. By using Eqs.(19) and (20) in Ref.\cite{4-Higgs-nnlo}, we obtain that the NLO and NNLO QCD $K$-factors for $pp \to gg \to HH+X$ in the SM at the $14~{\rm TeV}$ LHC ($33~{\rm TeV}$ HE-LHC) are $1.873$ ($1.697$) and $2.272$ ($2.029$), respectively. It shows that the NLO and NNLO QCD corrections enhance the LO cross section for the gluon-gluon fusion Higgs pair production channel in the SM significantly. When we take into account these higher order QCD corrections, the relative RS effect will be suppressed by the corresponding QCD $K$-factor approximately. For example, the relative RS effects at the $14~ {\rm TeV}$ LHC and $33~ {\rm TeV}$ HE-LHC for $M_{HH}^{cut}=1000~{\rm GeV}$ are reduced to about $179\%$ and $466\%$, respectively, if the NLO QCD corrections to $pp \to gg \to HH+X$ in the SM are taken into consideration.
\begin{table}[htbp]
  \centering
    \begin{tabular}{c|cccc}
    \multicolumn{5}{c}{~~~$\sqrt{S}=14~{\rm TeV}$} \\
\hline
      $M_{HH}^{cut}$ [GeV] & ~~$\sigma_{SM}$ $[fb]$~~ & ~$\sigma_{RS}^{0}$ $[fb]$~ & ~~$\sigma_{RS}$ $[fb]$~~ & ~~$\delta$ $[\%]$~~ \\
\hline \hline
    300   & 14.619  & 14.960  & 15.088  & 3.21  \\
    400   & 10.127  & 10.468  & 10.595  & 4.62  \\
    500   & 4.571  & 4.912  & 5.038  & 10.22  \\
    600   & 2.014  & 2.355  & 2.479  & 23.09  \\
    700   & 0.938  & 1.279  & 1.403  & 49.6  \\
    800   & 0.467  & 0.808  & 0.931  & 99.4  \\
    900   & 0.247  & 0.588  & 0.711  & 188  \\
    \textbf{1000}  & \textbf{0.138}  & \textbf{0.478}  & \textbf{0.601}  & \textbf{336}  \\
\hline
    \multicolumn{5}{c}{} \\
    \multicolumn{5}{c}{~~~$\sqrt{S}=33~{\rm TeV}$} \\
\hline
      $M_{HH}^{cut}$ [GeV] & ~~$\sigma_{SM}$ $[fb]$~~ & ~$\sigma_{RS}^{0}$ $[fb]$~ & ~~$\sigma_{RS}$ $[fb]$~~ & ~~$\delta$ $[\%]$~~ \\
\hline \hline
    300   & 90.30  & 100.57  & 103.77  & 14.92  \\
    400   & 66.24  & 76.51  & 79.71  & 20.33  \\
    500   & 33.35  & 43.61  & 46.81  & 40.36  \\
    600   & 16.44  & 26.70  & 29.89  & 81.81  \\
    700   & 8.54  & 18.81  & 21.99  & 157  \\
    800   & 4.72  & 14.99  & 18.17  & 285  \\
    900   & 2.77  & 13.03  & 16.21  & 485  \\
    \textbf{1000}  & \textbf{1.70}  & \textbf{11.97}  & \textbf{15.15}  & \textbf{791}  \\
\hline
    \end{tabular}%
\caption{\label{tab2} The integrated cross sections ($\sigma_{SM}$, $\sigma_{RS}^{0}$ and $\sigma_{RS}$)
and corresponding relative RS effects for the $pp \to HH+X$ process at the $14~{\rm TeV}$ LHC and $33~{\rm TeV}$ HE-LHC with $\mu=\mu_1$ for some typical values of $M_{HH}^{cut}$.}
\end{table}

\par
Table \ref{tab3} is to show the dependence of the integrated cross section on the factorization/ renormalization scale. We list the integrated cross sections for $pp \to HH + X$ in both the SM and RS model ($\sigma_{SM}$, $\sigma_{RS}^0$ and $\sigma_{RS}$) at the $14~ {\rm TeV}$ LHC and $33~ {\rm TeV}$ HE-LHC with $\mu= \mu_0 /2$, $\mu_0$, $2\mu_0$, $\mu_1 /2$, $\mu_1$ and $2\mu_1$, separately. To estimate the theoretical uncertainty from the factorization/renormalization scale, we define the relative scale uncertainty as $\eta = \frac{\sigma(\mu=\mu_0/2) - \sigma(\mu=2\mu_0)}{\sigma(\mu=\mu_0)}$. Then we obtain $\eta = 72\%$, $55\%$ in the SM and $\eta = 15\%$, $11\%$ in the RS model at the $14~ {\rm TeV}$ LHC and $33~ {\rm TeV}$ HE-LHC, respectively. It shows that the integrated cross sections in both the SM and RS model are sensitive to the factorization/renormalization scale. However, it is more appropriate to take the dynamic factorization/renormalization scale $\mu=\mu_1$ in the calculation of the $pp \to HH + X$ process in the RS model, because the contribution to the double Higgs production at a high energy hadron collider is mainly from the KK graviton resonance when $\sqrt{\hat{s}}$ is larger than KK graviton mass. In the following calculation we take $\mu =\mu_1$.
\begin{table}[htbp]
  \centering
    \begin{tabular}{c|cccccc}
\multicolumn{7}{c}{$\sqrt{S}=14~{\rm TeV}$} \\
\hline
    ~~~~$\mu$~~~    & $\mu_0 /2$ & $\mu_0$ & $2\mu_0$ & $\mu_1 /2$ & $\mu_1$ & $2\mu_1$ \\
\hline \hline
    ~$\sigma_{SM}$ $[fb]$~  & 0.52970(9) & 0.36887(7) & 0.26494(4) & 0.18143(3) & 0.13764(3) & 0.10639(2)\\
    $\sigma_{RS}^0$ $[fb]$  & 1.41120(9)  & 1.07971(7) & 0.85104(5) & 0.54349(4) & 0.47839(3) & 0.40561(3)\\
    $\sigma_{RS}$ $[fb]$    & 0.9125(8)   & 0.8554(6)   & 0.7854(4)   & 0.6722(4) & 0.6014(3) & 0.5416(3)\\
\hline
\multicolumn{7}{c}{} \\
\multicolumn{7}{c}{$\sqrt{S}=33~{\rm TeV}$} \\
\hline
    $\mu$ & $\mu_0 /2$ & $\mu_0$ & $2\mu_0$ & $\mu_1 /2$ & $\mu_1$ & $2\mu_1$  \\
\hline \hline
    $\sigma_{SM}$ $[fb]$  & 4.9700(8) & 3.7642(7) & 2.9063(5) & 2.1213(4) & 1.7011(3) & 1.3819(3)\\
    $\sigma_{RS}^0$ $[fb]$& 24.504(2) & 20.825(2) & 17.951(2) & 13.444(1) & 11.965(1) & 10.740(1)\\
    $\sigma_{RS}$ $[fb]$  & 20.18(3)  & 19.11(3)   & 18.09(2)  & 16.16(2) & 15.15(2) & 14.24(2)\\
\hline
    \end{tabular}%
  \caption{ The integrated cross sections for the $pp \to HH+X$ process in the SM and RS model at the $14~{\rm TeV}$ LHC and $33~{\rm TeV}$ HE-LHC for some typical values of the factorization/renormalization scale.}
  \label{tab3}
\end{table}

\par
\subsection{Kinematic distributions }
\par
\begin{figure}[!htbp]
   \begin{center}
     \includegraphics[scale=0.27]{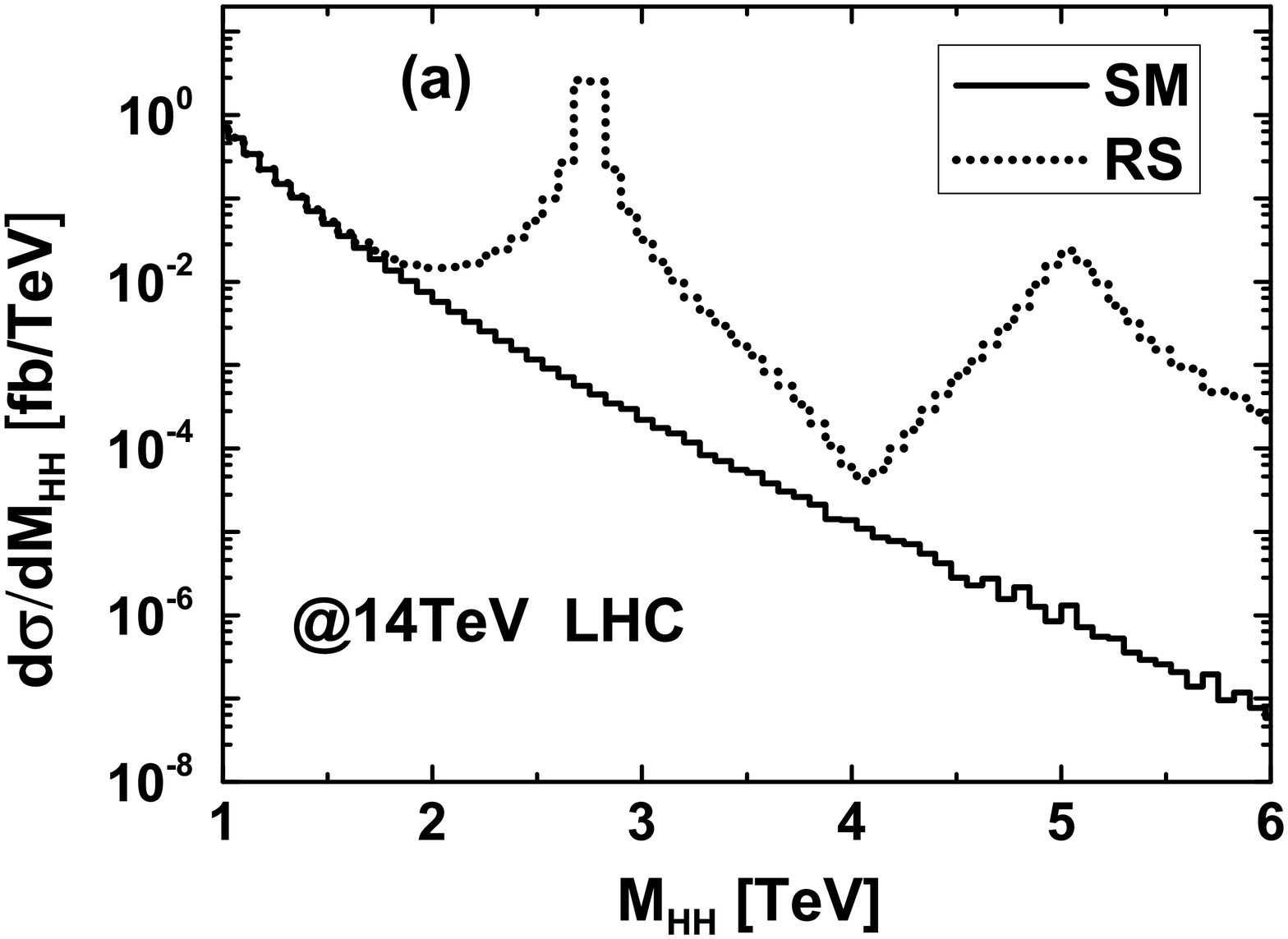}
     \includegraphics[scale=0.27]{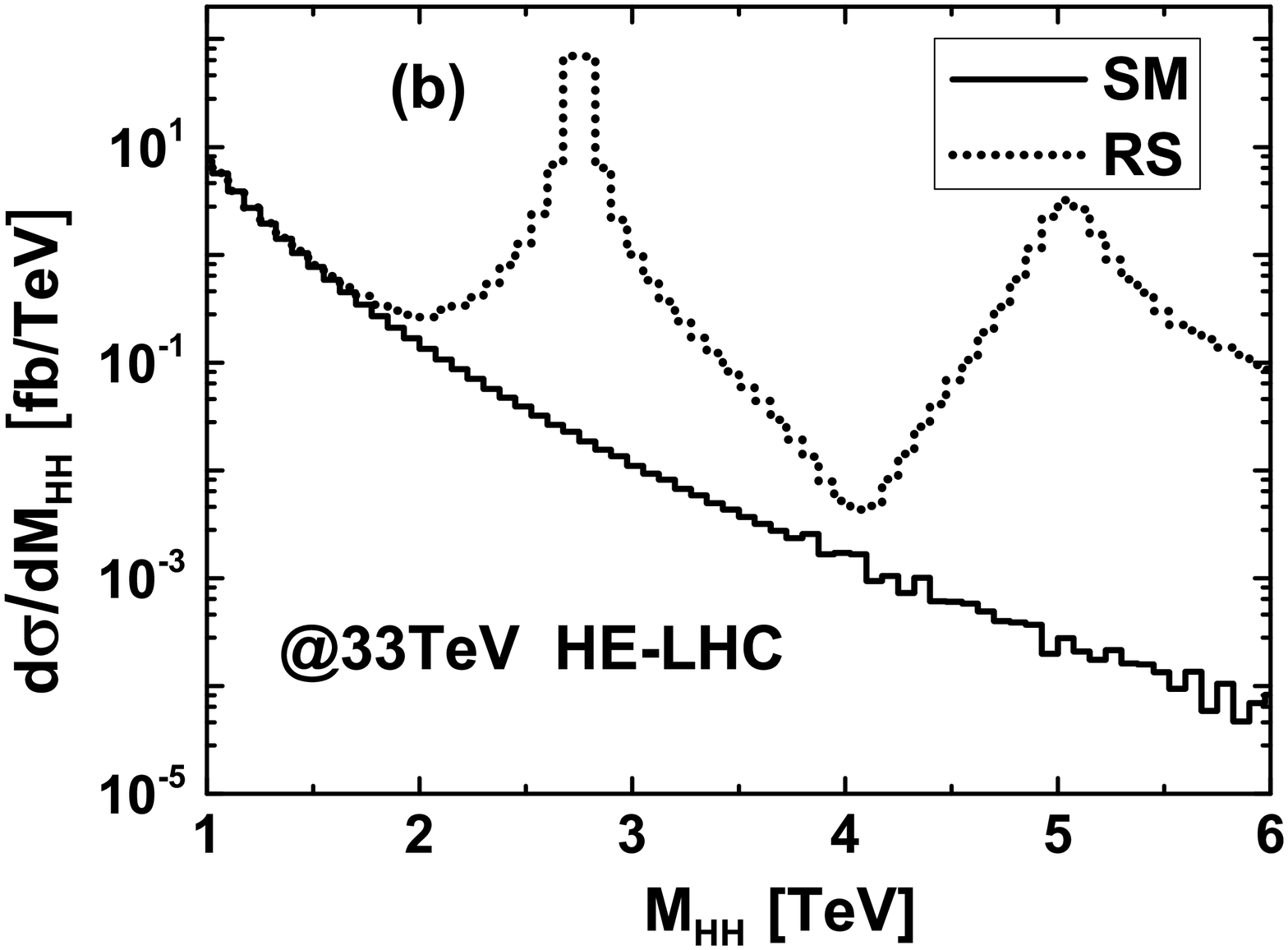}
     \caption{\label{fig5} The Higgs pair invariant mass distributions for the $pp \to HH + X$ process in the SM and RS model at (a) $14~{\rm TeV}$ LHC and (b) $33~{\rm TeV}$ HE-LHC.}
   \end{center}
\end{figure}

Now we turn to the RS effect on the kinematic distributions of final products. In Figs.\ref{fig5}(a) and (b) we present the Higgs pair invariant mass distributions for $pp \to HH + X$ in the SM and RS model at the $14~{\rm TeV}$ LHC and $33~{\rm TeV}$ HE-LHC, respectively. As shown in the figures the SM-like contribution to the $M_{HH}$ distribution is dominant in low invariant mass region, i.e., $1000~{\rm GeV} < M_{HH} < 1500~{\rm GeV}$. With the increment of $M_{HH}$, the SM-like contribution decreases significantly and the sensitivity to KK graviton resonance becomes more obvious. This behavior is also shown in Table \ref{tab2}. From the two figures we can see that there are two peaks around the masses of the first two KK gravitons, i.e., $M_{HH} \sim M_1=2.75~{\rm TeV}$ and $M_{HH} \sim M_2=5.04~{\rm TeV}$, for the $M_{HH}$ distribution in the RS model. It indicates that the resonances of the lightest two KK gravitons contribute to the double Higgs boson production significantly. Therefore, it is possible to apply a mass window cut on the Higgs pair invariant mass to enhance the RS effect.

\par
\begin{figure}[!htbp]
   \begin{center}
      \includegraphics[scale=0.27]{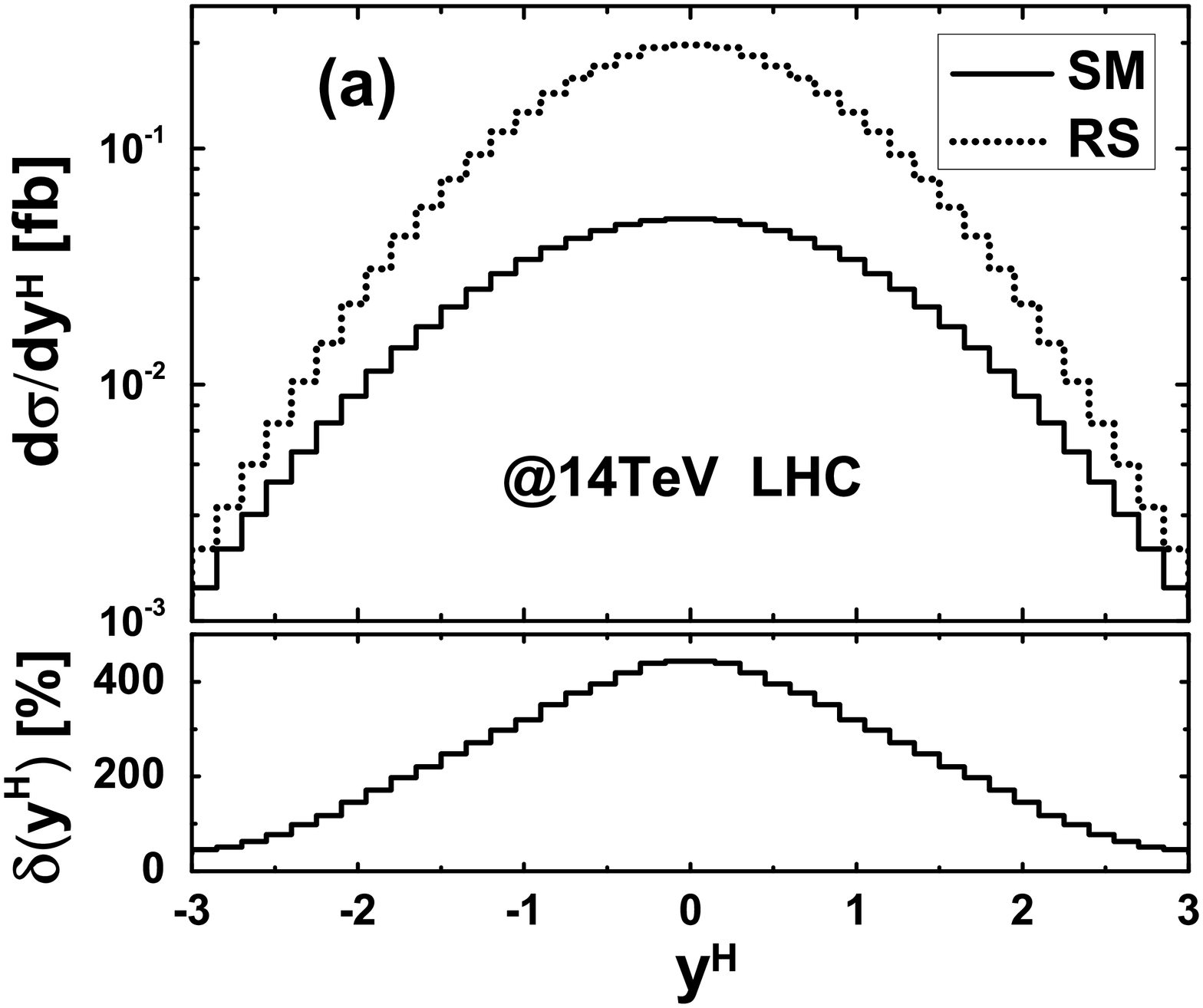}
      \includegraphics[scale=0.27]{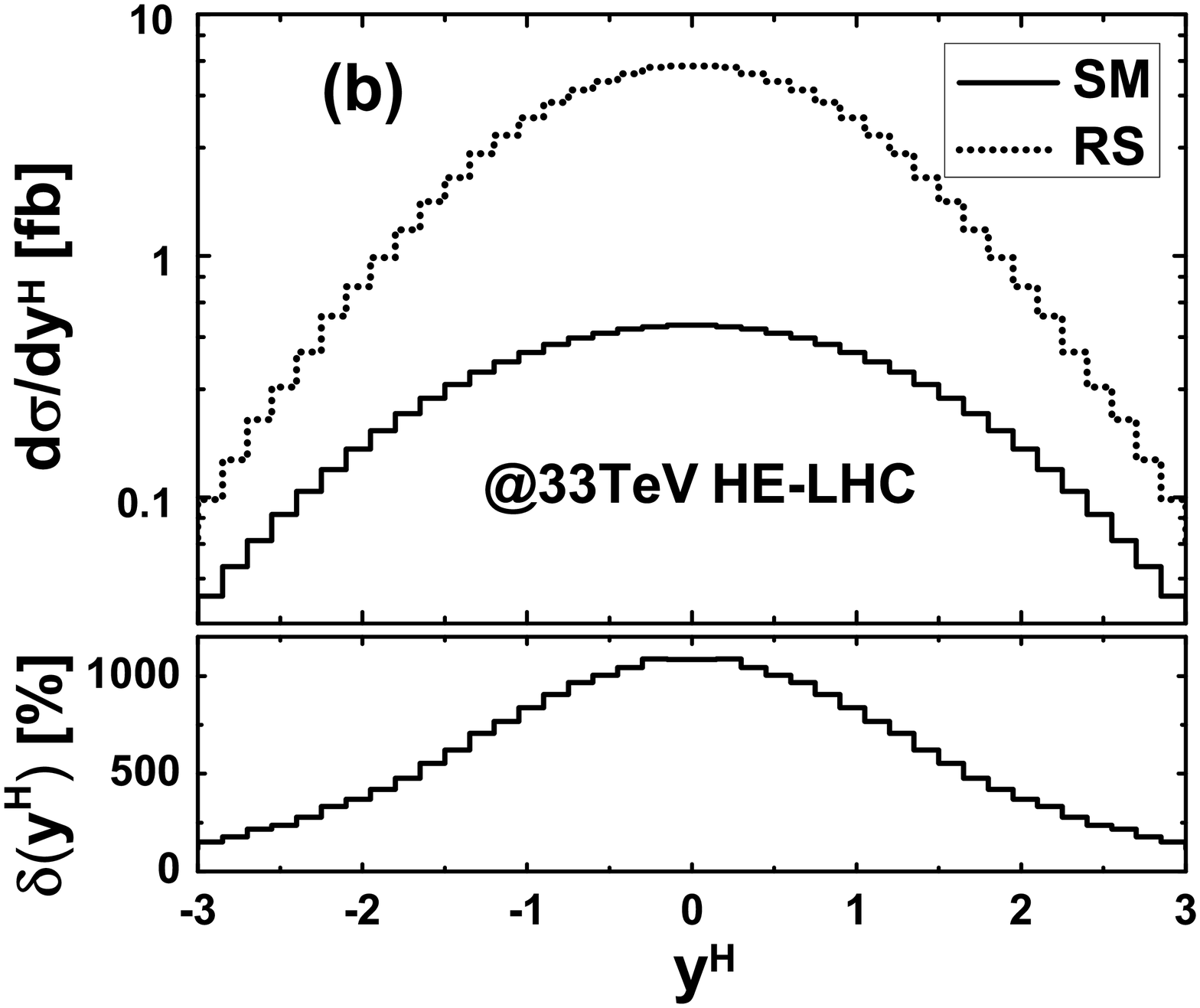}
      \caption{\label{fig6} The final Higgs boson rapidity distributions for the $pp \to HH + X$ process in the SM and RS model at (a) $14~{\rm TeV}$ LHC and (b) $33~{\rm TeV}$ HE-LHC.}
   \end{center}
\end{figure}
\begin{figure}[!htbp]
   \begin{center}
      \includegraphics[scale=0.27]{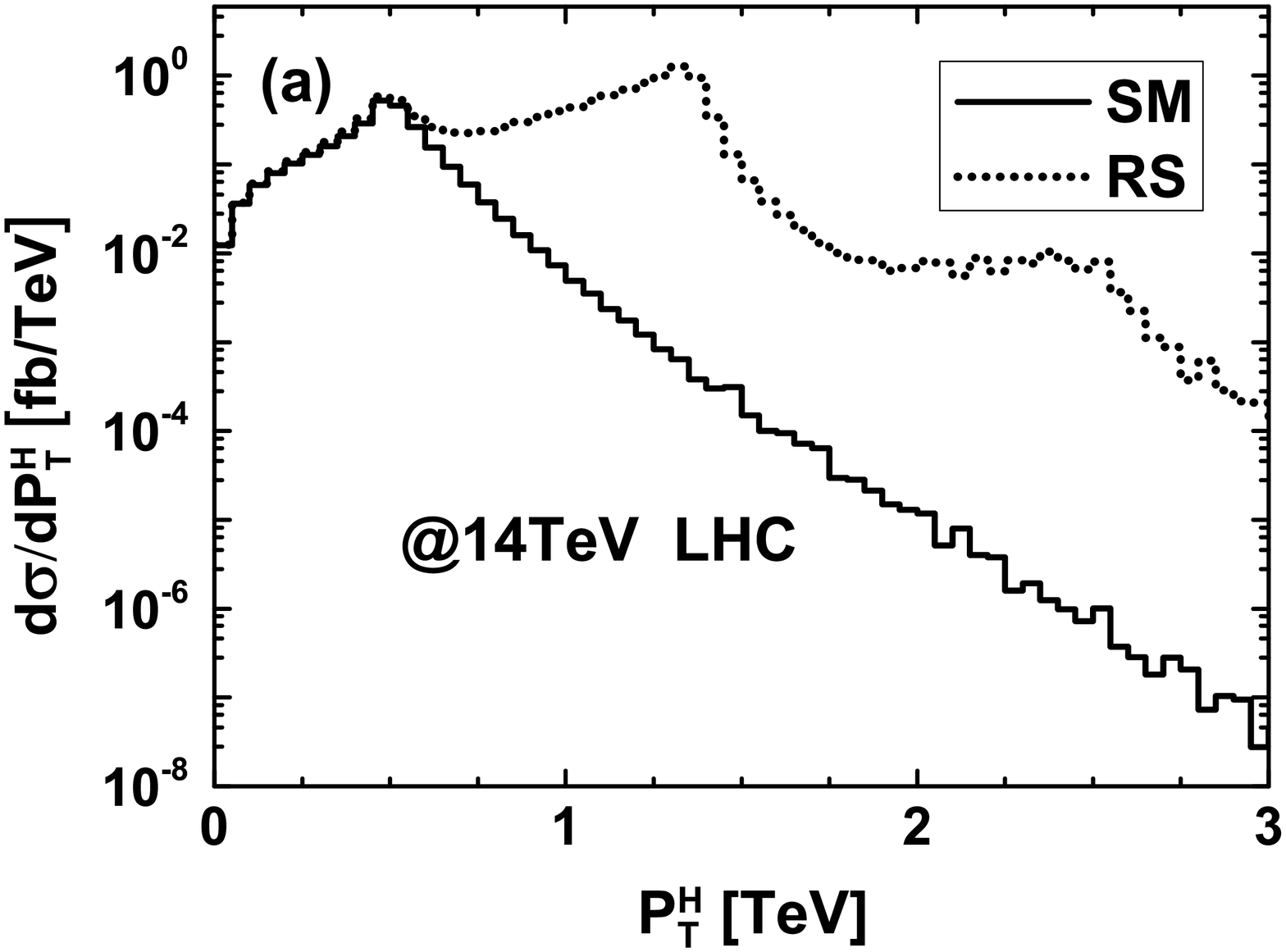}
      \includegraphics[scale=0.27]{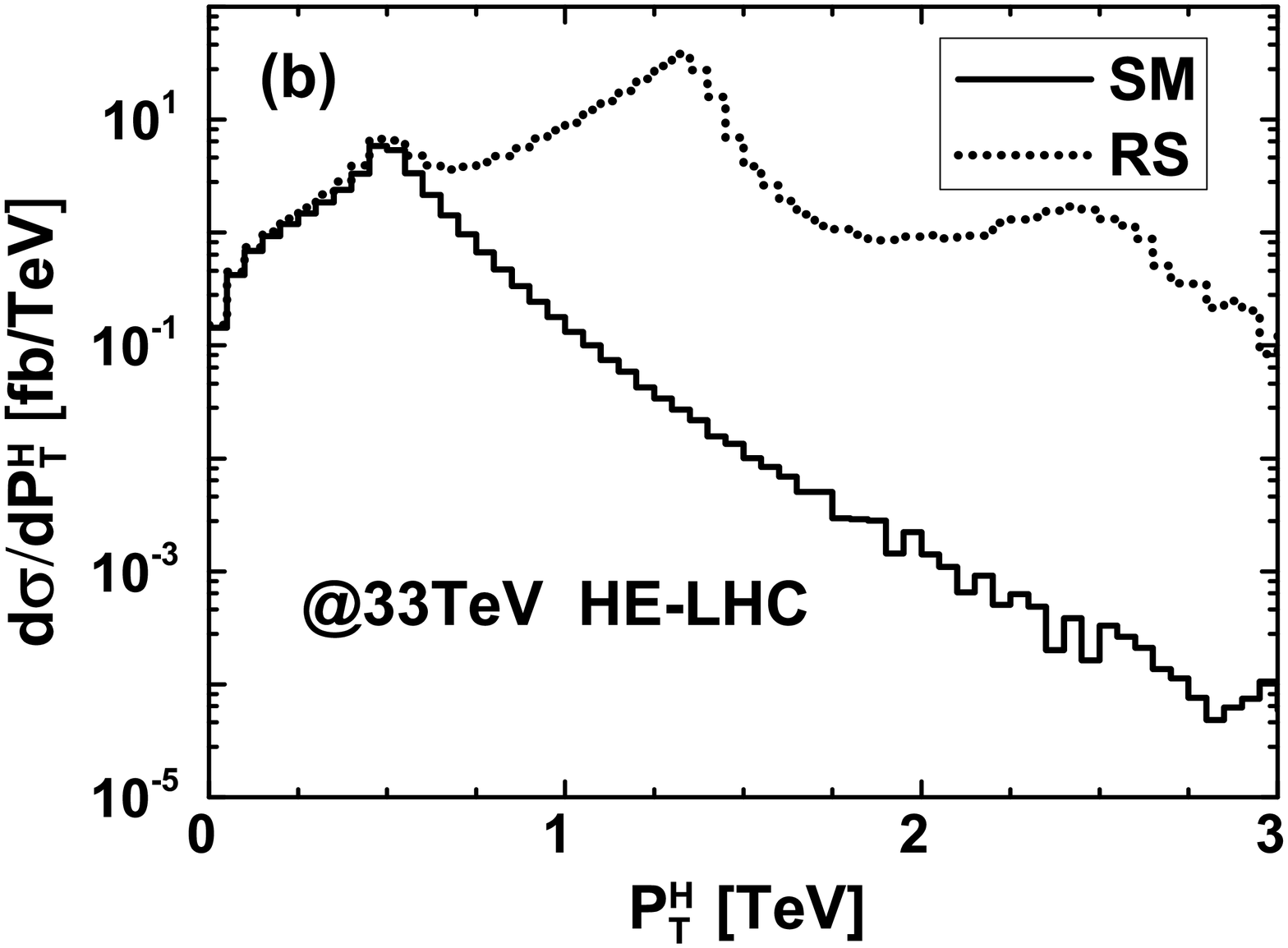}
      \caption{\label{fig7} The final Higgs boson transverse momentum distributions for the $pp \to HH + X$ process in the SM and RS model at (a) $14~{\rm TeV}$ LHC and (b) $33~{\rm TeV}$ HE-LHC.}
   \end{center}
\end{figure}

In Figs.\ref{fig6}(a) and (b) we depict the rapidity distributions of final Higgs bosons for $pp \to HH + X$ in both the SM and RS model at the $14~ {\rm TeV}$ LHC and $33~ {\rm TeV}$ HE-LHC, separately. Since there are two identical Higgs bosons in the final state, we take the rapidities of  both final Higgs bosons as entries in the histograms \footnote{The histograms for rapidity as well as transverse momentum distributions of final Higgs bosons (Figs.\ref{fig6} and \ref{fig7}) should be divided by 2 for normalization to total cross section.}. We define the relative RS effect on the rapidity distribution of final Higgs bosons as $\delta(y^{H}) = \left(\frac{d\sigma_{RS}}{dy^{H}}-\frac{d\sigma_{SM}}{dy^{H}}\right)\Big/\frac{d\sigma_{SM}}{dy^{H}}$. From the figures we see clearly that the Higgs rapidity distributions in both the SM and RS model and the relative RS effect concentrate in the central rapidity region, and reach their maxima at $y^H = 0$. We can read out from the figures that $\delta(y^H = 0) \sim 443\%$ and $1083\%$ at the $14~ {\rm TeV}$ LHC and $33~ {\rm TeV}$ HE-LHC, respectively.

\par
We also present the transverse momentum distributions of final Higgs bosons for $pp \to HH + X$ at the $14~{\rm TeV}$ LHC and $33~{\rm TeV}$ HE-LHC in Figs.\ref{fig7}(a) and (b), separately. Similar to the data taking method in Figs.\ref{fig6}(a,b), we pick the transverse momenta of both final Higgs bosons and fill them in the histograms in Figs.\ref{fig7}(a) and (b). We can see from Figs.\ref{fig7}(a,b) that there are two peaks on each curve for $d \sigma_{RS}/dp_T^H$, one is located at $p_T^H \sim \frac{1}{2} M_1 = 1.375~ {\rm TeV}$ and the other, which actually looks like a bulge, is implicitly in the vicinity of $p_T^H \sim \frac{1}{2} M_2 = 2.52~ {\rm TeV}$. As the increment of the $pp$ colliding energy from $14~{\rm TeV}$ to $33~{\rm TeV}$, the second bulge stands out and the RS effect is enhanced in high $p_T^H$ region. From all the distributions in Figs.\ref{fig5}(a,b), Figs.\ref{fig6}(a,b) and Figs.\ref{fig7}(a,b), we can see clearly that the RS effect is significant in some kinematic regions, such as $M_{HH} \sim M_{1,2}$, $y^H \sim 0$, and $p_T^H \sim \frac{1}{2} M_{1,2}$.

\par
\subsection{RS model parameter dependence}
\par
\begin{figure}[!htbp]
\begin{center}
   \includegraphics[scale=0.27]{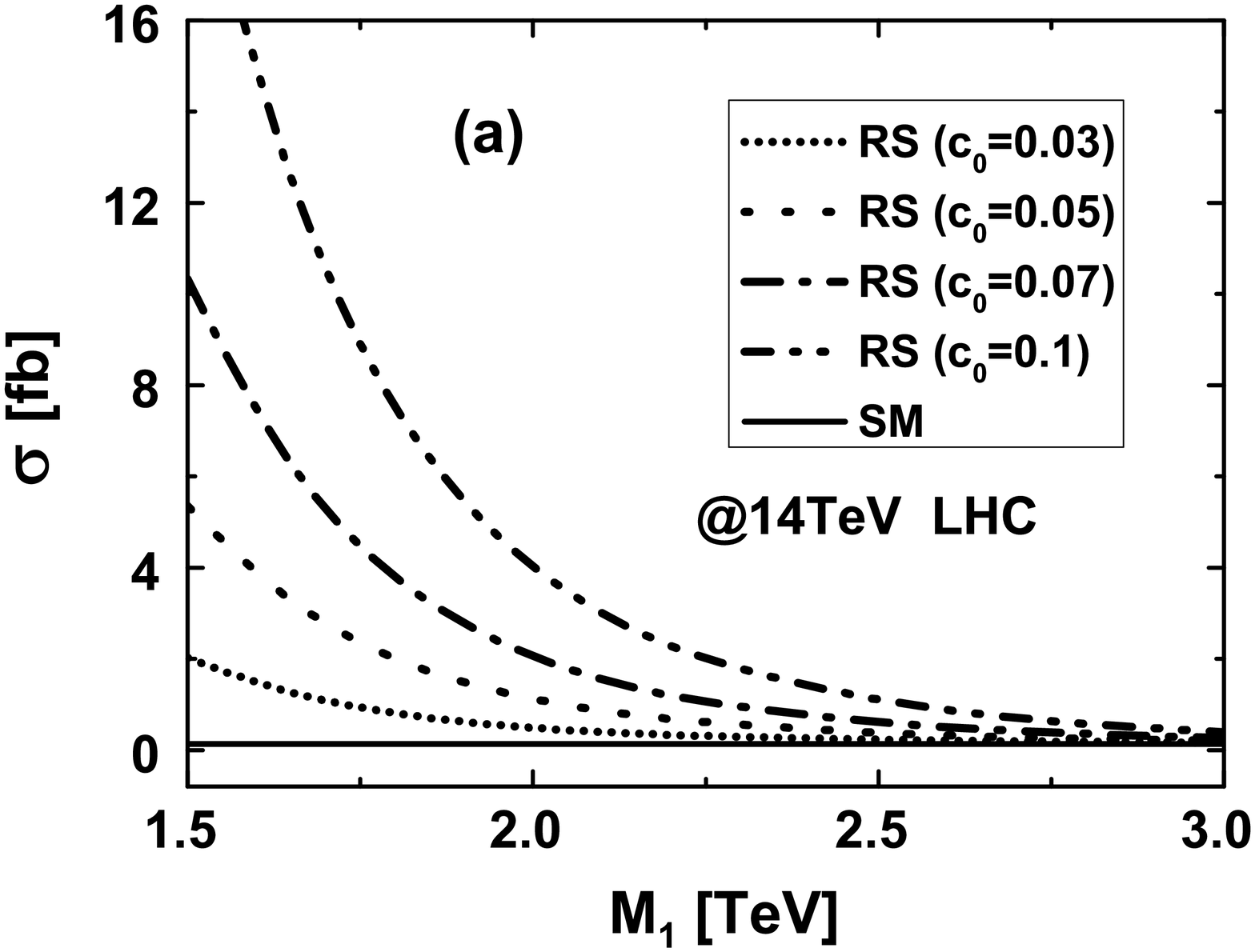}
   \includegraphics[scale=0.27]{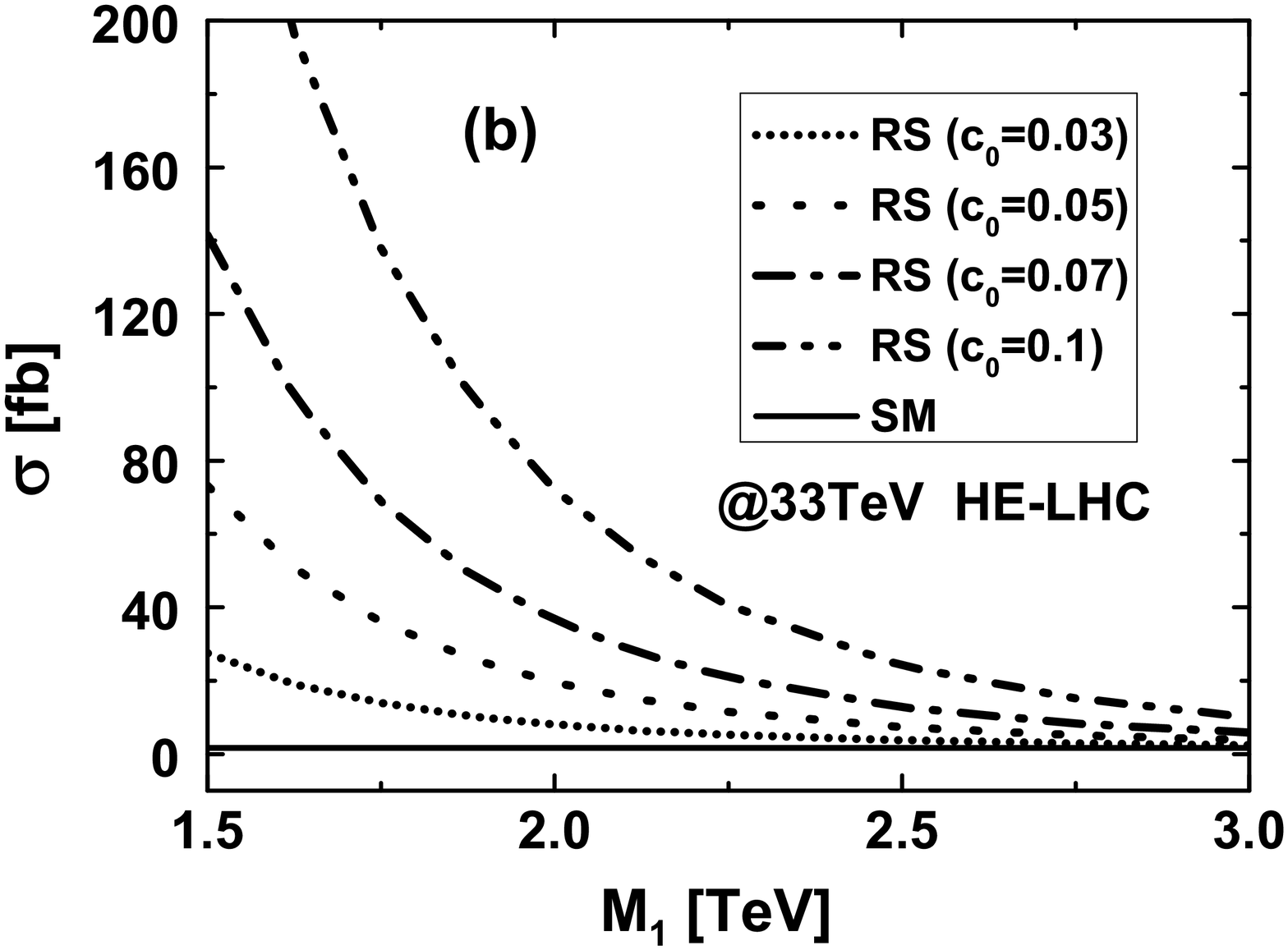}
	\caption{\label{fig8} The integrated cross sections for the $pp \to HH+X$ process as functions of the first KK graviton mass $M_1$ with $c_0=0.03$, $0.05$, $0.07$ and $0.1$ at (a) $14~{\rm TeV}$ LHC and (b) $33~{\rm TeV}$ HE-LHC. }
\end{center}
\end{figure}

We investigate the dependence of the integrated cross section for $pp \to HH + X$ on the RS model parameters $M_1$ and $c_{0}$. In our discussion we take $c_0=0.03$, $0.05$, $0.07$, $0.1$, and vary $M_{1}$ in the range of $[1.5,~3.0]~{\rm TeV}$. The integrated cross sections at the $14~ {\rm TeV}$ LHC and $33~ {\rm TeV}$ HE-LHC are depicted in Figs.\ref{fig8}(a) and (b), respectively. The horizontal full lines correspond to the cross sections in the SM. From the figures we see that the integrated cross section in the RS model decreases with the increment of the first KK graviton mass $M_1$. For instance, we can read out from the curves for $c_0=0.03$ that the integrated cross sections in the RS model can reach $2.02~fb$ and $27.44~fb$ at $M_1 = 1.5~ {\rm TeV}$, and decrease to $0.160~fb$ and $2.46~fb$ at $M_1 = 3.0~ {\rm TeV}$ which are almost the same as the corresponding SM predictions, at the $14~ {\rm TeV}$ LHC and $33~ {\rm TeV}$ HE-LHC, respectively. We also see that the integrated cross section in the RS model is reduced obviously with the decrement of the effective coupling $c_0$. When $M_1 = 2.75~ {\rm TeV}$, $c_0=0.03$, $0.05$, $0.07$ and $0.1$, the relative RS effects for the $pp \to HH+X$ process at the $14~ {\rm TeV}$ LHC ($33~ {\rm TeV}$ HE-LHC) are $30.5\%$ ($72.0\%$), $84.8\%$ ($199\%$), $166\%$ ($391\%$) and $336\%$ ($791\%$), correspondingly.

\par
\subsection{Double Higgs boson production with semileptonic decay}
\par
\begin{figure}[!htbp]
   \begin{center}
      \includegraphics[scale=0.27]{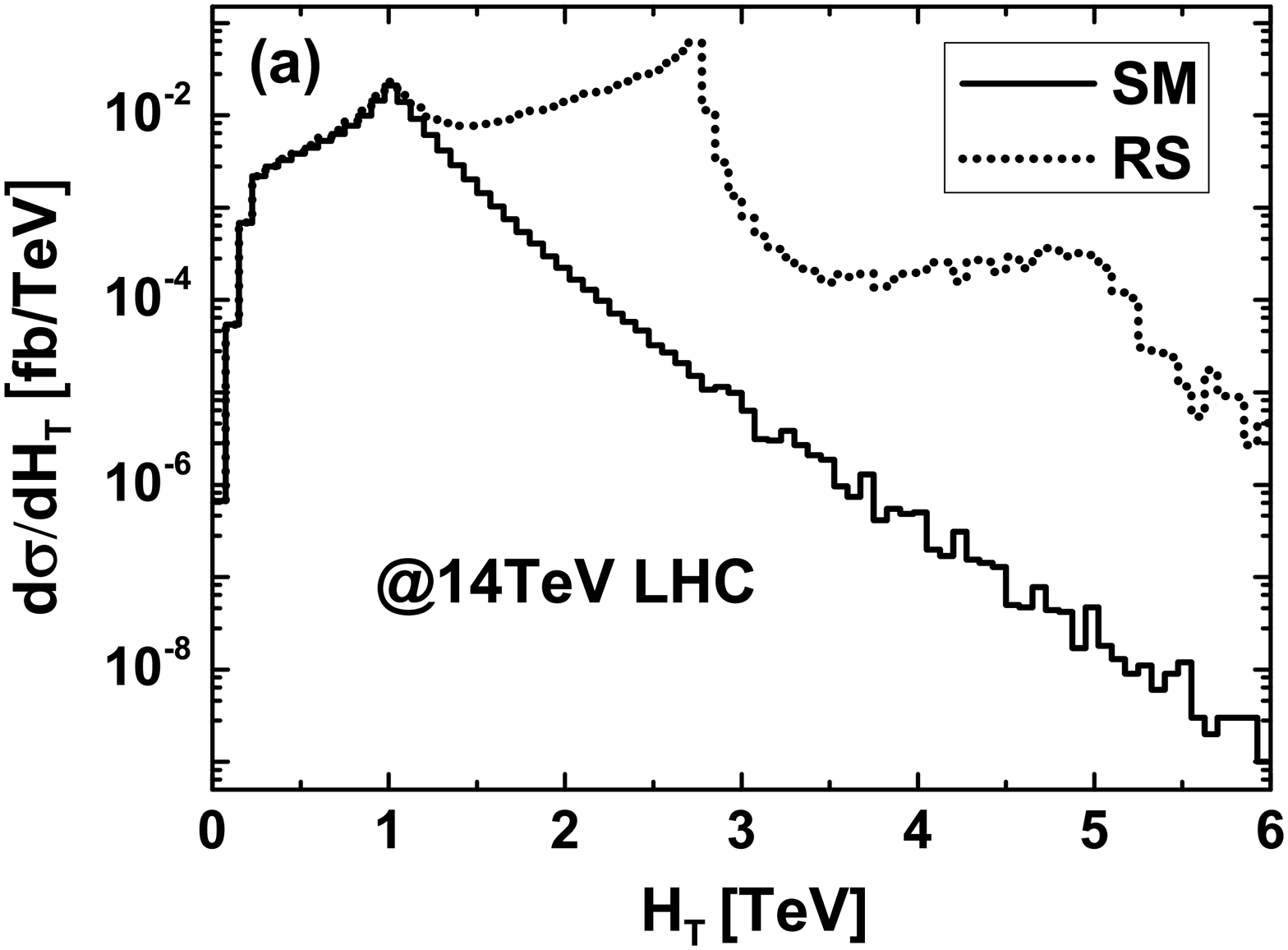}
      \includegraphics[scale=0.27]{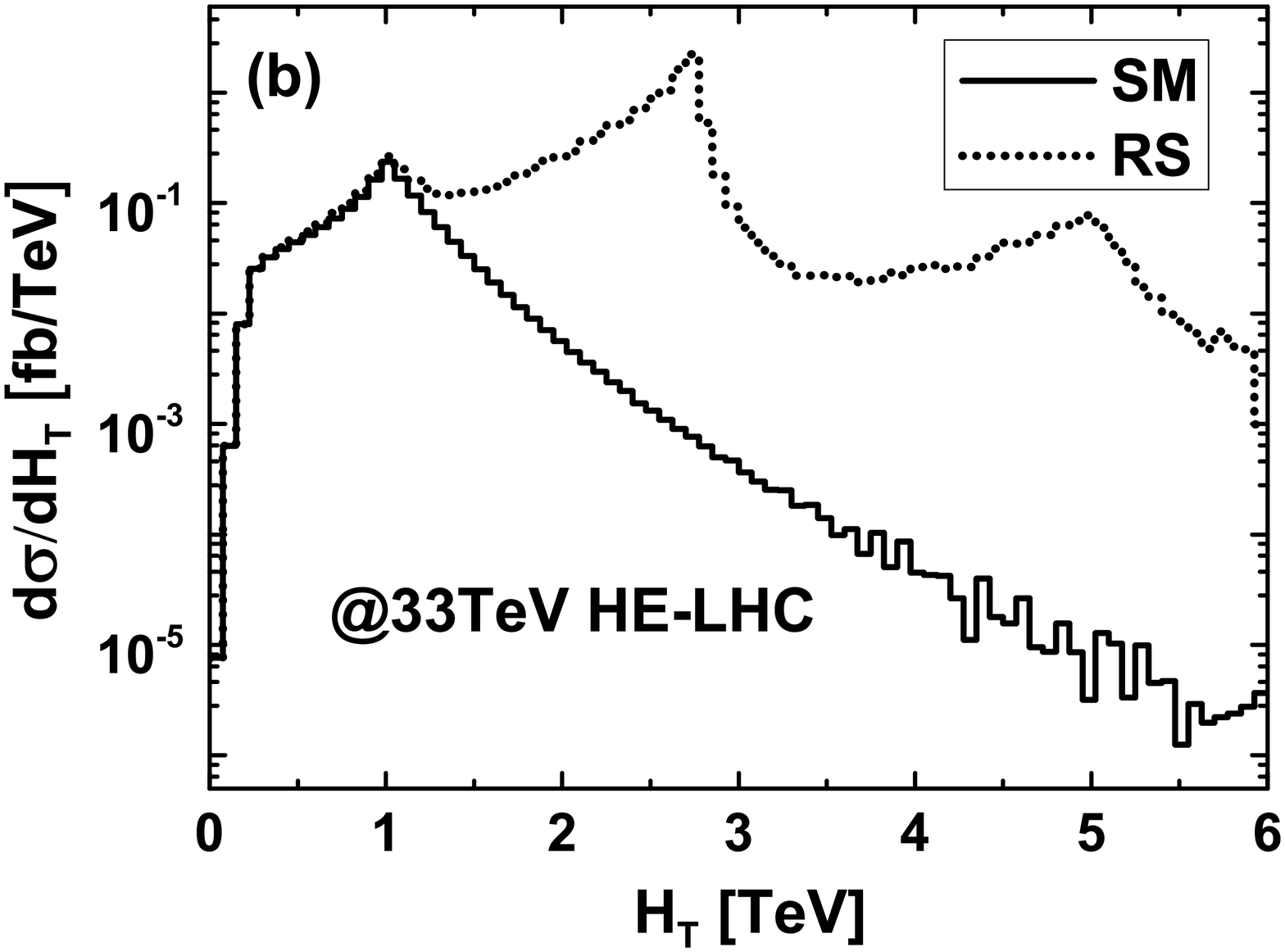}
      \caption{\label{fig9} The $H_T$ distributions for the $pp \to HH \to b \bar{b} \tau^+ \tau^- + X$ process in the SM and RS model at (a) $14~{\rm TeV}$ LHC and (b) $33~{\rm TeV}$ HE-LHC.  }
   \end{center}
\end{figure}

We consider the Higgs pair production with subsequent Higgs boson decays by employing the narrow width approximation. The Higgs decay channels are chosen as: one Higgs boson decays to $b\bar{b}$ and the other to $\tau^+\tau^-$. The branch ratios are obtained as $Br(H\to b\bar{b})=59.29\%$ and $Br(H\to \tau^+\tau^-)=5.782\%$ by using the HDECAY program. In Figs.\ref{fig9}(a) and (b), we demonstrate the $H_T$ distributions for the signal process $pp \to HH \to b\bar{b} \tau^+\tau^- + X$ in the SM and RS model at the $14~ {\rm TeV}$ LHC and $33~ {\rm TeV}$ HE-LHC, respectively, where $H_T=\sum_{i}\left|\vec{p}_T (i)\right|$ is the scalar sum of the transverse momenta of final quarks and leptons, i.e., $b$, $\bar{b}$, $\tau^+$ and $\tau^-$. From the two figures we find that the $H_T$ distribution in the RS model has two peaks at $H_T \sim M_1$ and $H_T \sim M_2$, and behaves similarly to the $M_{HH}$ distribution in the RS model. We also see that the $H_T$ distribution in the SM declines seriously with the increment of $H_T$ in the region of $H_T > 1000~{\rm GeV}$.

\par
\begin{figure}[!htbp]
   \begin{center}
      \includegraphics[scale=0.27]{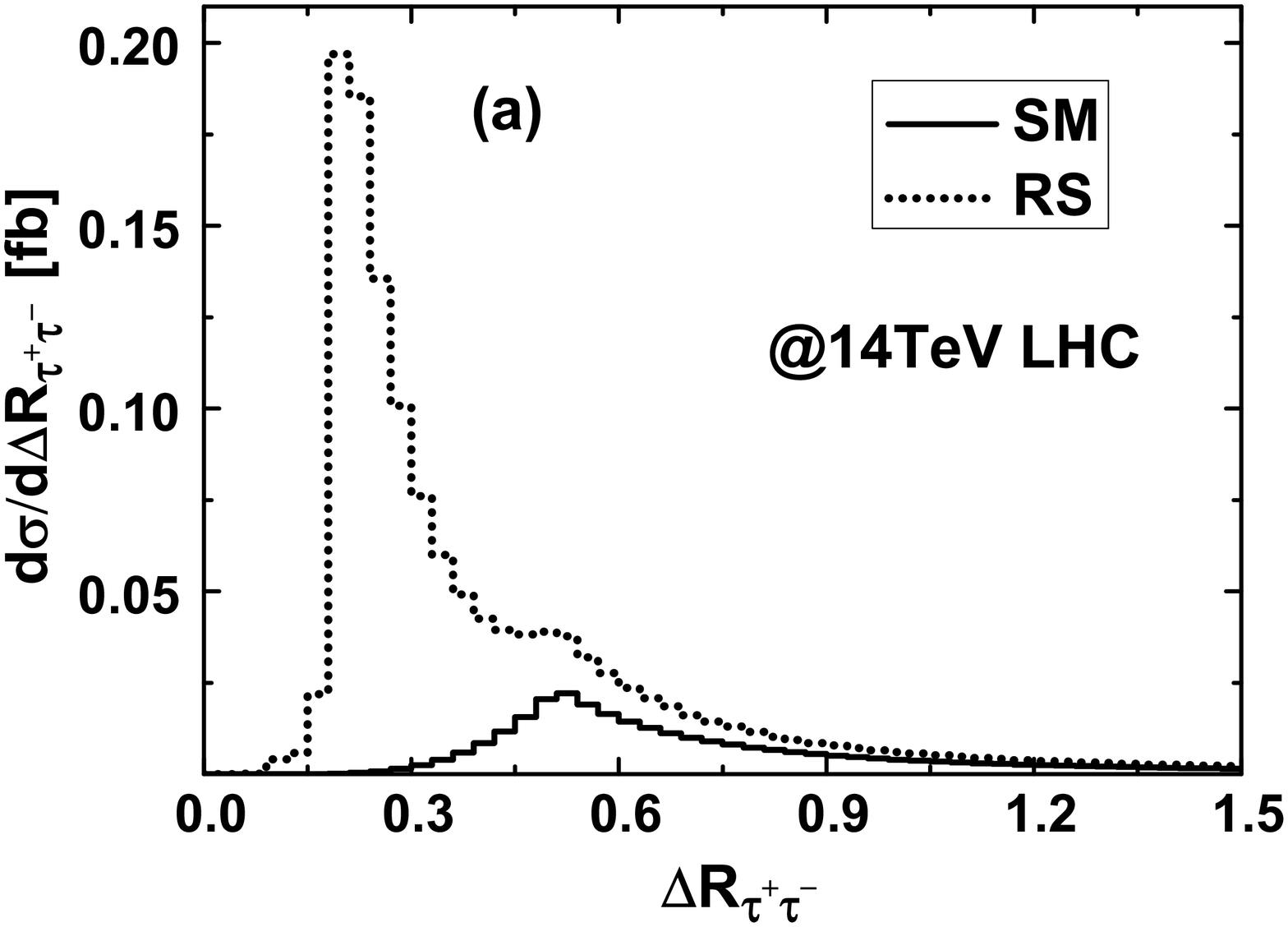}
      \includegraphics[scale=0.27]{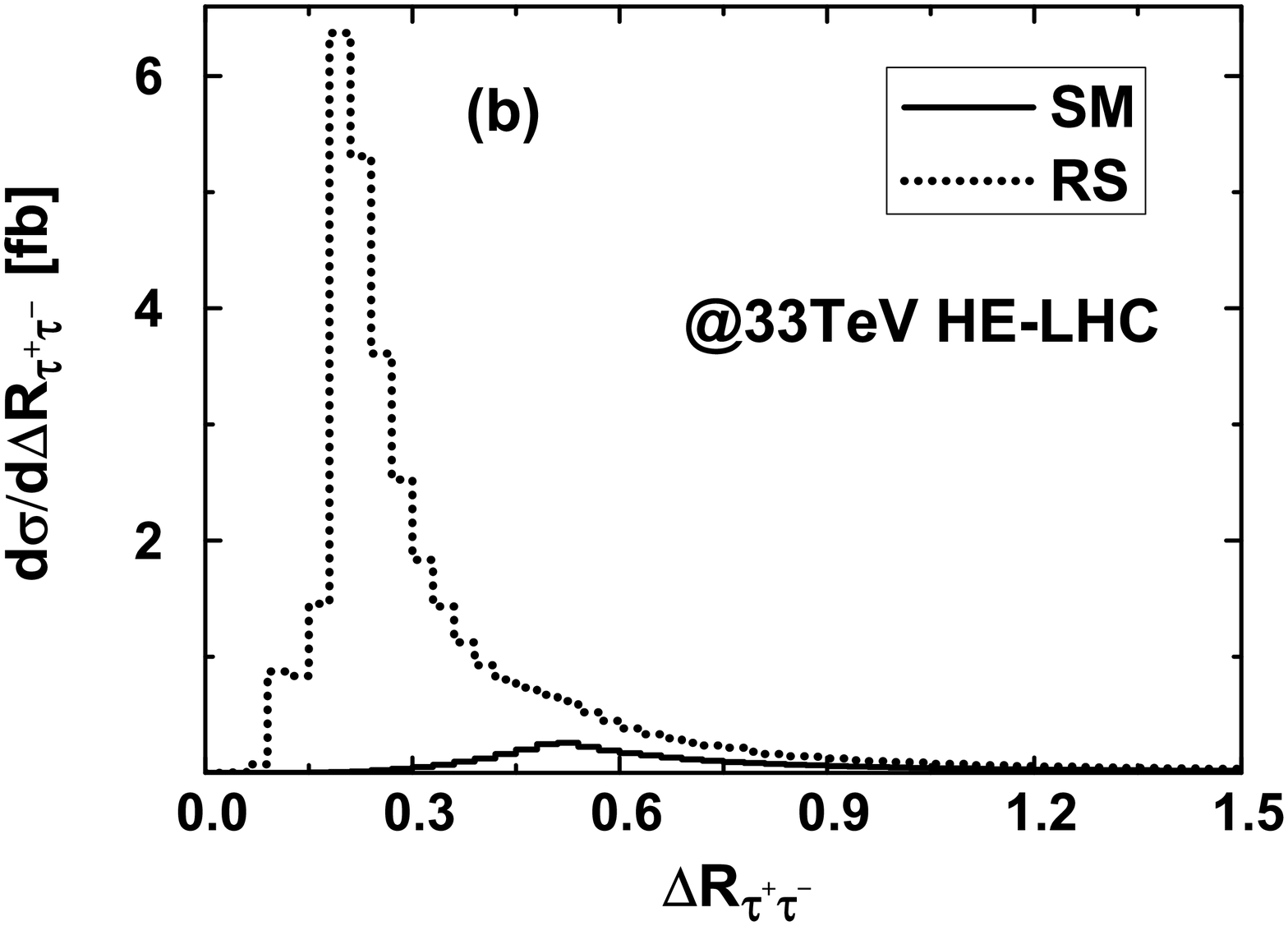}
      \caption{\label{fig10} The $\Delta R_{\tau^+\tau^-}$ distributions for the $pp \to HH \to b\bar{b}\tau^+\tau^- + X$ process in the SM and RS model at (a) $14~{\rm TeV}$ LHC and (b) $33~{\rm TeV}$ HE-LHC. }
   \end{center}
\end{figure}

The separation between $\tau^+$ and $\tau^-$ on the rapidity-azimuthal-angle plane is defined as $\Delta R_{\tau^+\tau^-}=\sqrt{\Delta y^2+\Delta \phi^2}$, where $\Delta y$ and $\Delta \phi$ are the differences of rapidity and azimuthal angle between $\tau^+$ and $\tau^-$. We depict the $\Delta R_{\tau^+\tau^-}$ distributions for $pp \to HH \to b\bar{b} \tau^+\tau^- + X$ in the SM and RS model at the $14~{\rm TeV}$ LHC and $33~{\rm TeV}$ HE-LHC in Figs.\ref{fig10}(a) and (b), separately. From the figures we see that the RS effect is quite significant in the region of $\Delta R_{\tau^+\tau^-}< 0.4$ and reaches its maximum at $\Delta R_{\tau^+\tau^-} \sim 0.2$, but is rather small and can be neglected when $\Delta R_{\tau^+\tau^-}>0.9$. This characteristic is due to the fact that the RS effect mainly comes from the resonant KK graviton production, in which the final Higgs bosons are energetic and therefore the separation between the sequentially produced $\tau^+$ and $\tau^-$ is small. It is foreseeable that the distribution of the separation between $b$ and $\bar{b}$ is almost the analogue as between $\tau^+$ and $\tau^-$, therefore is not given in this paper.

\par
\begin{figure}[!htbp]
   \begin{center}
      \includegraphics[scale=0.27]{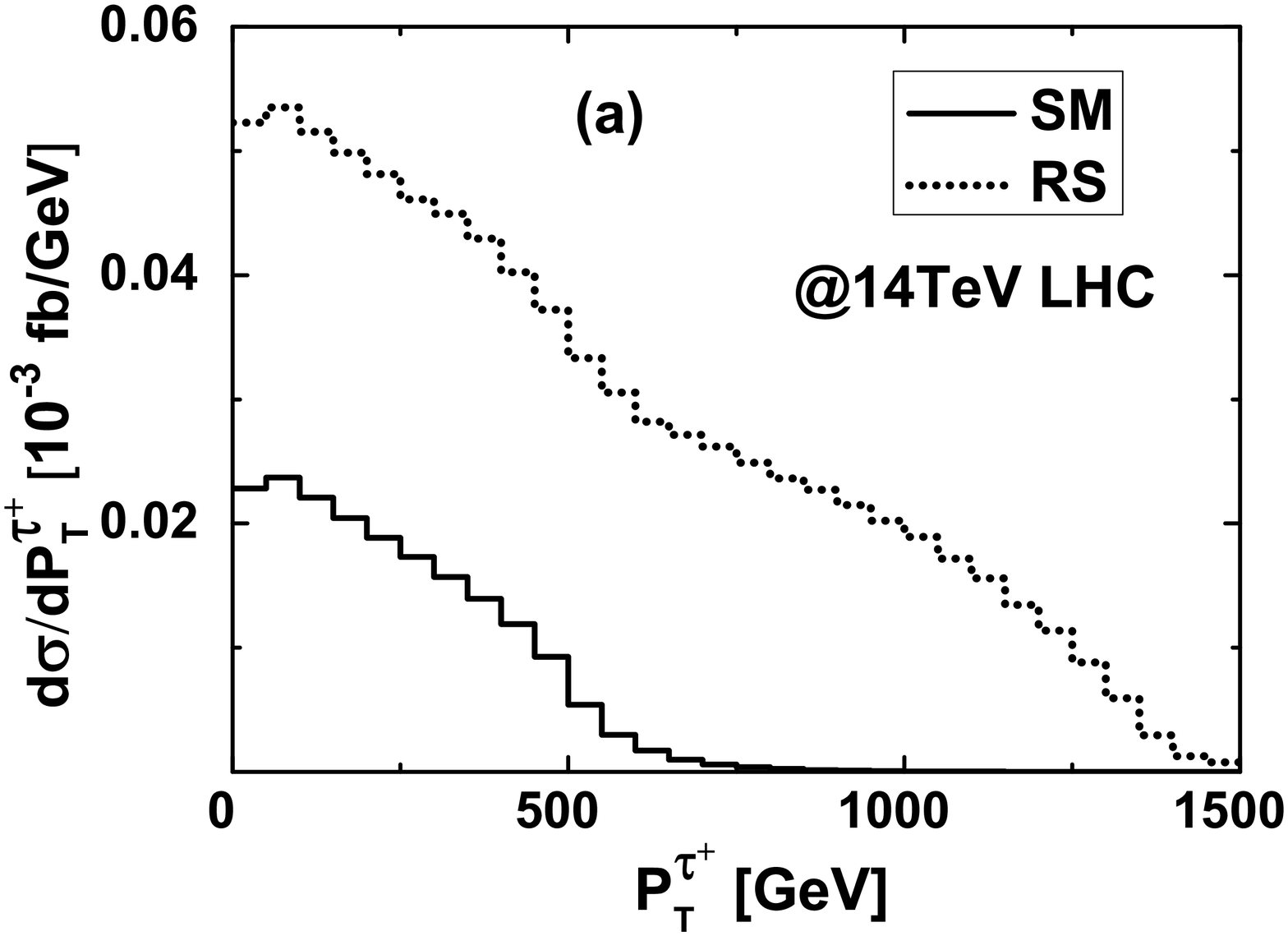}
      \includegraphics[scale=0.27]{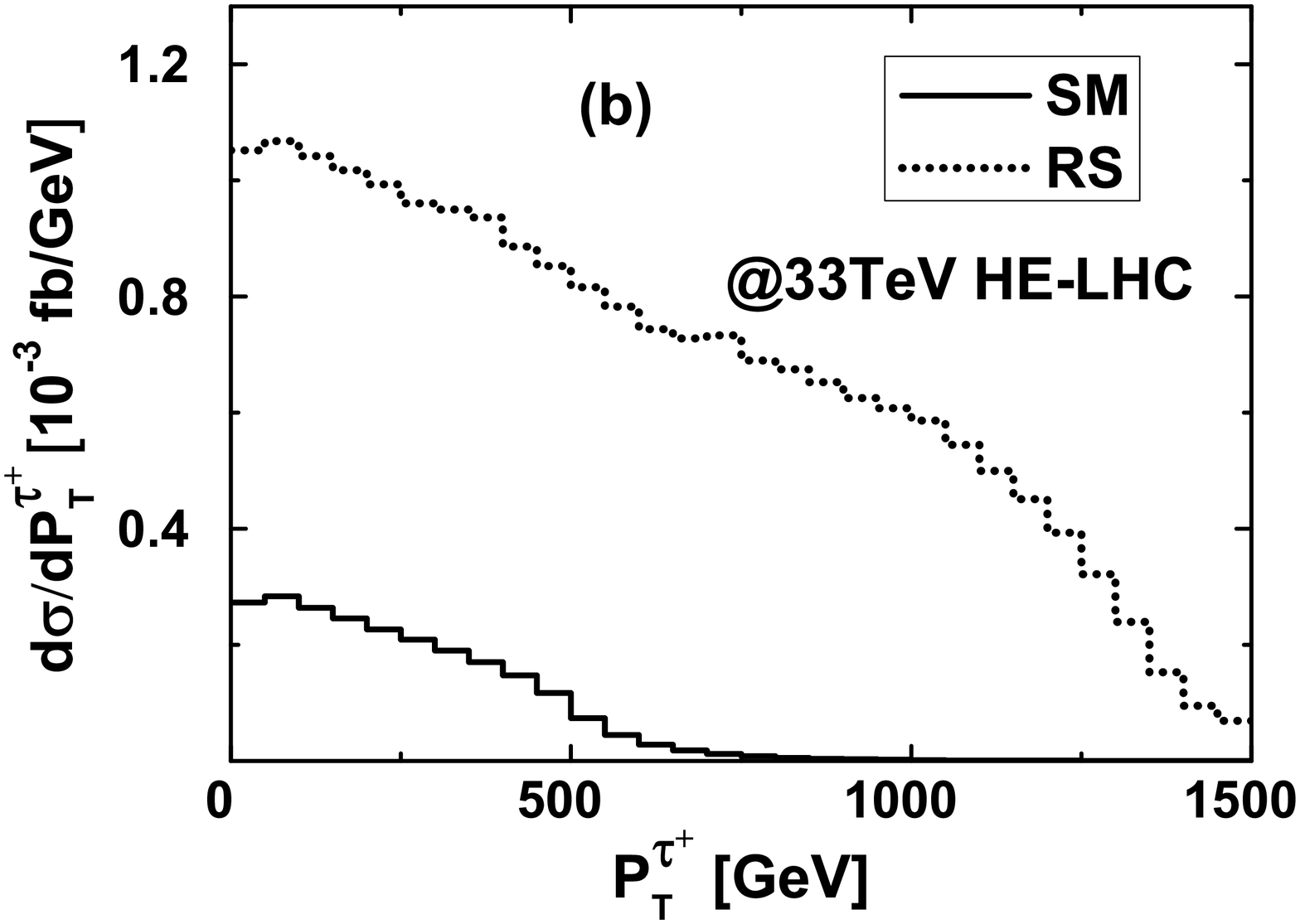}
      \caption{\label{fig11} The transverse momentum distributions of $\tau^+$ for the $pp \to HH \to b\bar{b}\tau^+\tau^-+X$ process in the SM and RS model at (a) $14~{\rm TeV}$ LHC and (b) $33~{\rm TeV}$ HE-LHC. }
   \end{center}
\end{figure}

We also plot the transverse momentum distributions of final $\tau^+$ for $pp \to HH \to b\bar{b} \tau^+\tau^- + X$ in the SM and RS model at the $14~{\rm TeV}$ LHC and $33~{\rm TeV}$ HE-LHC in Figs.\ref{fig11}(a) and (b), separately. Unlike the $p_T^H$ distribution where the RS effect is obvious in high $p_T^H$ region, the RS effect on the $p_T^{\tau^+}$ distribution is significant in the whole plotted $p_T^{\tau^+}$ region, and becomes larger and larger with the increment of $p_T^{\tau^+}$ in the region of $p_T^{\tau^+} < 1350~ {\rm GeV}$. The SM-like contribution is negligible when $p_T^{\tau^+} > 750~{\rm GeV}$.

\par
In detecting the RS effect on the $pp\to HH \to b\bar{b}\tau^+\tau^-+X$ process at the LHC or HE-LHC, we should suppress the accompanied SM backgrounds. The main backgrounds to the RS effect on Higgs pair production are the $pp \to ZZ \to b\bar{b}\tau^+\tau^- + X$, $pp \to ZH \to b\bar{b}\tau^+\tau^- + X$, $pp \to t\bar{t} \to b\bar{b}\tau^+\tau^-\nu_{\tau}\bar{\nu}_{\tau} + X$ processes, as well as the SM contribution to the $pp \to HH \to b\bar{b}\tau^+\tau^- + X$ process. The mass difference between $Z$ boson and Higgs boson is about $35~{\rm GeV}$, therefore almost all the $b\bar{b}\tau^+\tau^-$ events from $ZZ$ and $ZH$ productions can be excluded by applying the invariant mass cuts of $|M_{b\bar{b}}-M_H| \leq 20~{\rm GeV}$ and $|M_{\tau^+\tau^-}-M_H| \leq 20~{\rm GeV}$ simultaneously. For $t\bar{t}$ production with subsequent decays $t \to W^+b \to \tau^+ \nu_{\tau} b$ and $\bar{t} \to W^-\bar{b} \to \tau^- \bar{\nu}_{\tau} \bar{b}$, we apply $p_T^{miss} < 50~{\rm GeV}$ on the missing transverse momentum in the final state to suppress this background. In Table \ref{tabx} we list the signal event numbers ($S$), significances ($\frac{S}{\sqrt{S+B}}$) and relative RS effects for $pp \to HH \to b\bar{b}\tau^+\tau^- + X$ at the $14~ {\rm TeV}$ LHC and $33~ {\rm TeV}$ HE-LHC with integrated luminosities of ${\cal L}=600$ and $1500~fb^{-1}$. For $M_{b\bar{b}\tau^+\tau^-}^{cut}=0$, the relative RS effects are $3.15\%$ and $14.75\%$ and the signal significances are about $4.7~(7.4)$ and $13.1~(20.8)$ for ${\cal L}=600~fb^{-1}$ ($1500~fb^{-1}$) at the $14~ {\rm TeV}$ LHC and $33~ {\rm TeV}$ HE-LHC, respectively. With the increment of $M_{b\bar{b}\tau^+\tau^-}^{cut}$, the signal event number declines seriously while the significance raises at first and then declines. However, the relative RS effect increases with the increment of $M_{b\bar{b}\tau^+\tau^-}^{cut}$. As the increment of $M_{b\bar{b}\tau^+\tau^-}^{cut}$ to $500~{\rm GeV}$, the signal event number drops to $207~(518)$ and $1926~(4814)$ for ${\cal L}=600~fb^{-1}$ ($1500~fb^{-1}$) at the $14~ {\rm TeV}$ LHC and $33~ {\rm TeV}$ HE-LHC, separately, but the RS effect could be remained at a visible level and the signal significance is also prominent. If we take a more stringent constraint of $M_{b\bar{b}\tau^+\tau^-} > 1000~{\rm GeV}$, which is not the optimum invariant mass cut for signal significance, the relative RS effects can reach about $336\%$ and $791\%$ at the $14~ {\rm TeV}$ LHC and $33~ {\rm TeV}$ HE-LHC, respectively. Therefore, we conclude that the double Higgs production is a promising process to explore RS effect at the LHC and HE-LHC.
\begin{table}[htbp]
  \small
  \centering
    \begin{tabular}{c|c|c|c|c|c}
    \multicolumn{6}{c}{$\sqrt{S}=14~{\rm TeV}$} \\
    \hline
    \multirow{2}*{$M_{b\bar{b}\tau^+\tau^-}^{cut}$[GeV]}  &\multicolumn{2}{c|}{$S$} & \multicolumn{2}{c|}{$\frac{S}{\sqrt{S+B}}$}&\multirow{2}*{~~~~~$\delta$ $[\%]$~~~~~} \\
    \cline{2-5}
    & ${\cal L}=600~fb^{-1}$ & ${\cal L}=1500~fb^{-1}$  & ${\cal L}=600~fb^{-1}$  & ${\cal L}=1500~fb^{-1}$   &  \\
    \hline
    0     & 631   & 1577  & 4.7   & 7.4   & 3.15  \\
    300   & 621   & 1552  & 6.4   & 10.0  & 3.21  \\
    400   & 436   & 1090  & 13.6  & 21.5  & 4.62  \\
    \textbf{500}   & \textbf{207}   & \textbf{518}   & \textbf{12.9}  & \textbf{20.3}  & \textbf{10.22}  \\
    600   & 102   & 255   & 9.7   & 15.4  & 23.09  \\
    700   & 58    & 144   & 7.5   & 11.9  & 49.6  \\
    800   & 38    & 96    & 6.2   & 9.7   & 99.4  \\
    900   & 29    & 73    & 5.4   & 8.5   & 188  \\
    \textbf{1000}  & \textbf{25}    & \textbf{62}    & \textbf{5.0}   & \textbf{7.9}   & \textbf{336}  \\
    \hline
    \multicolumn{6}{c}{} \\
    \multicolumn{6}{c}{$\sqrt{S}=33~{\rm TeV}$} \\
    \hline
    \multirow{2}*{$M_{b\bar{b}\tau^+\tau^-}^{cut}$[GeV]}  &\multicolumn{2}{c|}{$S$} & \multicolumn{2}{c|}{$\frac{S}{\sqrt{S+B}}$}&\multirow{2}*{$\delta$ $[\%]$}\\
    \cline{2-5}
    & ${\cal L}=600~fb^{-1}$ & ${\cal L}=1500~fb^{-1}$  & ${\cal L}=600~fb^{-1}$  & ${\cal L}=1500~fb^{-1}$   &  \\
   \hline
    0     & 4319  & 10797  & 13.1  & 20.8  & 14.75  \\
    300   & 4269  & 10672  & 17.4  & 27.5  & 14.92  \\
    400   & 3279  & 8198  & 37.8  & 59.8  & 20.33  \\
    \textbf{500}   & \textbf{1926}  & \textbf{4814}  & \textbf{39.9}  & \textbf{63.0}  & \textbf{40.36}  \\
    600   & 1230  & 3074  & 34.4  & 54.4  & 81.81  \\
    700   & 905   & 2262  & 30.0  & 47.4  & 157  \\
    800   & 747   & 1869  & 27.3  & 43.2  & 285  \\
    900   & 667   & 1667  & 25.8  & 40.8  & 485  \\
    \textbf{1000}  & \textbf{623}   & \textbf{1557}  & \textbf{25.0}  & \textbf{39.5}  & \textbf{791}  \\
    \hline
    \end{tabular}%
  \caption{The signal event numbers, significances  and relative RS effects for $pp \rightarrow HH \rightarrow b\bar{b}\tau^+\tau^- + X$ at the $14~ {\rm TeV}$ LHC and $33~ {\rm TeV}$ HE-LHC with ${\cal L}=600$ and $1500~fb^{-1}$. }
  \label{tabx}%
\end{table}

\vskip 5mm
\section{Summary}
\par
In this work we study the possible Randall-Sundrum effect on the double Higgs boson production at the $14~{\rm TeV}$ LHC and $33~{\rm TeV}$ HE-LHC. We consider only the lightest two KK gravitons which provide the dominant contribution to the RS effect. We present the integrated cross sections and some kinematic distributions of final products in both the SM and RS model. The results show that the relative RS effect in the vicinities of $M_{HH} \sim M_{1}$, $M_{2}$ or in the central Higgs rapidity region is quite significant. We find that with the increment of $M_{HH}^{cut}$, the integrated cross section in the SM declines more quickly than that in the RS model and then the relative RS effect becomes significant. We conclude that with proper kinematic cuts the double Higgs boson production process is promising in detecting the RS effect. We also investigate the dependence of the integrated cross section on the RS model parameters $M_1$ and $c_0$, and find that the integrated cross section is reduced obviously with the increment of the first KK graviton mass $M_1$ or the decrement of the effective coupling $c_0$. We find that if we take $pp\to HH \to b\bar{b}\tau^+\tau^-+X$ as the signal process to probe RS effect, it is possible to extract the RS effect on the signal events from the heavy SM background by choosing proper kinematic cuts on final particles.

\vskip 5mm
\section{Acknowledgments}
This work was supported in part by the National Natural Science Foundation of China (Grants No.11275190, No.11375008, No.11375171, No.11405173) and the Fundamental Research Funds for the Central Universities (Grant No.WK2030040044).

\vskip 5mm
\noindent{\large\bf Appendix: The Relevant Feynman Rules}
\par
The relevant RS couplings \cite{16-RS-coupling} used in our calculation are listed below:
\begin{itemize}
\par
\item[(i)]
    $G_{KK}^{\mu\nu}(k_3)-H(k_1)-H(k_2)$ vertex:
\begin{eqnarray}
     -i\frac{1}{\Lambda_{\pi}}\left[\eta^{\mu\nu}(m_{H}^2+k_1\cdot k_2)-k_1^{\mu}k_2^{\nu}-
     k_1^{\nu}k_2^{\mu}\right]
\end{eqnarray}
\item[(ii)]
    $G_{KK}^{\mu\nu}(k_3)-\overline{\psi}(k_1)-\psi(k_2)$ vertex:
\begin{eqnarray}
     -i\frac{1}{4\Lambda_{\pi}}\left[\gamma^{\mu}(k_1+k_2)^{\nu}+\gamma^{\nu}(k_1+k_2)^{\mu}
     -2\eta^{\mu\nu}(\rlap/{k_1}+\rlap/{k_2}-2m_{\psi})\right]
\end{eqnarray}
\item[(iii)]
    $G_{KK}^{\mu\nu}(k_4)-\overline{\psi}(k_1)-\psi(k_2)-A^{a\rho}(k_3)$ vertex:
\begin{eqnarray}
     ig_s\frac{1}{2\Lambda_{\pi}}\left[\gamma^{\mu}\eta^{\nu\rho}+\gamma^{\nu}\eta^{\mu\rho}
     -2\gamma^{\rho}\eta^{\mu\nu}\right]T^{a}
\end{eqnarray}
\item[(iv)]
    $G_{KK}^{\mu\nu}(k_3)-A^{a\rho}(k_1)-A^{b\sigma}(k_2)$ vertex:
\begin{eqnarray}
     i\frac{2}{\Lambda_{\pi}}\delta^{ab}
     \left[(C^{\mu \nu \rho \sigma \tau \beta}
     -C^{\mu\nu\rho\beta\sigma\tau})
     k_{1\tau}k_{2\beta}+\frac{1}{\xi}
     E^{\mu\nu\rho\sigma}(k_1,k_2)\right]
\end{eqnarray}
\item[(v)]
    $G_{KK}^{\mu\nu}(k_4)-A^{a\rho}(k_1)-A^{b\sigma}(k_2)-A^{c\lambda}(k_3)$ vertex:
\begin{eqnarray}
     \frac{2}{\Lambda_{\pi}}g_s f^{abc}\left[(k_1-k_3)_{\tau}C^{\mu\nu\tau\sigma\rho\lambda}
     +(k_2-k_1)_{\tau}C^{\mu\nu\sigma\rho\tau\lambda}+(k_3-k_2)_{\tau}C^{\mu\nu\lambda\sigma\tau\rho}\right]
\end{eqnarray}
\item[(vi)]
    $G_{KK}^{\mu\nu}(k_5)-A^{a\rho}(k_1)-A^{b\sigma}(k_2)-A^{c\lambda}(k_3)-A^{d\delta}(k_4)$ vertex:
\begin{eqnarray}
     -i\frac{1}{\Lambda_{\pi}}g_s^2\left[f^{eac}f^{ebd}D^{\mu\nu\rho\sigma\lambda\delta}
     +f^{eab}f^{ecd}D^{\mu\nu\rho\lambda\sigma\delta}+f^{ead}f^{ebc}D^{\mu\nu\rho\sigma\delta\lambda}\right]
\end{eqnarray}
\item[(vii)]
    $G_{KK}^{\mu\nu}(k_3)-\overline{\eta}^{a}(k_1)-\eta^{b}(k_2)$ vertex:
\begin{eqnarray}
     -i\frac{2}{\Lambda_{\pi}}\delta^{ab}B^{\alpha\beta\mu\nu}k_{1\alpha}k_{2\beta}
\end{eqnarray}
\item[(viii)]
    $G_{KK}^{\mu\nu}(k_4)-\overline{\eta}^{a}(k_1)-\eta^{b}(k_2)-A^{c\rho}(k_3)$ vertex:
\begin{eqnarray}
     \frac{2}{\Lambda_{\pi}}g_sf^{abc}B^{\alpha\rho\mu\nu}k_{1\alpha}
\end{eqnarray}
\end{itemize}
where $G_{KK}^{\mu\nu}$, $H$, $\psi$, $A^{a\rho}$ and $\eta^{a}$ represent the fields of RS KK graviton, Higgs boson, fermion, gluon and ghost for gluon, respectively. In our calculation we adopt the Feynman gauge, i.e., $\xi=1$. The tensor coefficients $B^{\mu\nu\alpha\beta}$, $C^{\rho\sigma\mu\nu\alpha\beta}$, $D^{\mu\nu\rho\sigma\lambda\delta}$ and $E^{\mu\nu\rho\sigma}(k_1,k_2)$ are expressed as \cite{19-tensor}:
\begin{eqnarray}
B^{\mu \nu \alpha \beta} & = & \frac{1}{2}
      (
      \eta^{\mu \nu}\eta^{\alpha \beta}
      -\eta^{\mu \alpha}\eta^{\nu \beta}
      -\eta^{\mu \beta}\eta^{\nu \alpha}
      ),
       \nb \\
C^{\rho \sigma \mu \nu \alpha \beta} & = & \frac{1}{2}
      [
      \eta^{\rho \sigma}\eta^{\mu \nu}\eta^{\alpha \beta}
     -(
     \eta^{\rho \mu}\eta^{\sigma \nu}\eta^{\alpha \beta}
      +\eta^{\rho \nu}\eta^{\sigma \mu}\eta^{\alpha \beta}
      +\eta^{\rho \alpha}\eta^{\sigma \beta}\eta^{\mu \nu}
      +\eta^{\rho \beta}\eta^{\sigma \alpha}\eta^{\mu \nu}
      )
      ],
      \nb \\
D^{\mu \nu \rho \sigma \lambda \delta} &=&
     \eta^{\mu \nu}
     (
     \eta^{\rho \sigma} \eta^{\lambda \delta} - \eta^{\rho \delta} \eta^{\lambda \sigma}
     )
     +
     [
     \eta^{\mu \rho} \eta^{\nu \delta} \eta^{\lambda \sigma} +
     \eta^{\mu \lambda} \eta^{\nu \sigma} \eta^{\rho \delta} -
     \eta^{\mu \rho} \eta^{\nu \sigma} \eta^{\lambda \delta} -
     \eta^{\mu \lambda} \eta^{\nu \delta} \eta^{\rho \sigma} +
     (\mu \leftrightarrow \nu)
     ], \nb \\
E^{\mu \nu \rho \sigma}(k_{1},k_{2}) & = &
      \eta^{\mu \nu}
      (k_1^{\rho} k_1^{\sigma} + k_2^{\rho} k_2^{\sigma}
      + k_1^{\rho} k_2^{\sigma})
      - \left[\eta^{\nu \sigma} k_1^{\mu} k_1^{\rho}
      + \eta^{\nu \rho} k_2^{\mu} k_2^{\sigma} + (\mu \leftrightarrow \nu)\right].
\end{eqnarray}


\end{document}